\documentclass[a4paper,11pt]{article}

\usepackage{jinstpub} 

\usepackage[lofdepth,lotdepth]{subfig}
\usepackage{lineno}

\title{Characterization of a CdZnTe detector for a low-power CubeSat application}

\author[a,1]{Giulio Lucchetta,\note{Corresponding author.}}
\author[a]{Markus Ackermann,}
\author[a,b]{David Berge,}
\author[a]{Ingo Bloch,}
\author[a]{Rolf B{\"u}hler,}
\author[a,b]{Hermann Kolanoski,}
\author[a]{Wolfgang Lange}
\author[a]{and Francesco Zappon}

\affiliation[a]{Deutsches Elektronen-Synchrotron DESY,\\Platanenallee 6, 15738 Zeuthen, Germany}
\affiliation[b]{Institut f{\"u}r Physik, Humboldt-Universit{\"a}t zu Berlin,\\Newtonstra{\ss}e 15, 12489 Berlin, Germany}

\emailAdd{giulio.lucchetta@desy.de}

\abstract{We report spectral and imaging performance of a pixelated CdZnTe detector custom designed for the \emph{MeVCube} project: a small Compton telescope on a CubeSat platform. \emph{MeVCube} is expected to cover the energy range between $200\;\mathrm{keV}$ and $4\;\mathrm{MeV}$, with a sensitivity comparable to the one of the last generation of larger satellites. In order to achieve this goal, an energy resolution of few percent in full width at half maximum (FWHM) and a $3$-D spatial resolution of few millimeters for the individual detectors are needed. The severe power constraints present in small satellites require very low power read-out electronics for the detector. Our read-out is based on the VATA450.3 ASIC developed by \emph{Ideas}, with a power consumption of only $0.25\;\mathrm{mW/channel}$, which exhibits good performance in terms of dynamic range, noise and linearity. A $2.0\;\mathrm{cm} \times 2.0\;\mathrm{cm} \times 1.5\;\mathrm{cm}$ CdZnTe detector, with a custom $8 \times 8$ pixel anode structure read-out by a VATA450.3 ASIC, has been tested. A preliminary read-out system for the cathode, based on a discrete \emph{Amptek} A250F charge sensitive pre-amplifier and a DRS4 ASIC, has been implemented. An energy resolution around $3\%$ FWHM has been measured at a gamma energy of $662\;\mathrm{keV}$; at $200\;\mathrm{keV}$ the average energy resolution is $6.5\%$, decreasing to $\lesssim 2\%$ at energies above $1\;\mathrm{MeV}$. A $3$-D spatial resolution of $\approx 2\,\mathrm{mm}$ is achieved in each dimension.}


\begin{document}
\maketitle
\flushbottom

\section{Introduction}
\label{sec:introduction}
Over the last few decades, Cadmium Zinc Telluride (CdZnTe or CZT) semiconductor detectors have gained increasing interest for X-ray and gamma-ray applications \cite{Schlesinger_ReviewCZT,DelSordo_ReviewCZT}. Compared to other traditional semiconductor materials such as silicon (Si) and germanium (Ge), CdZnTe displays a higher atomic number and density, which translates into high radiation stopping power and detection efficiency. At the same time, the wider band gap and low leakage currents are favourable for low noise measurement and promise an excellent energy resolution performance at room temperature. In the current state of technology CdZnTe detectors can be produced with high quality up to few $\mathrm{cm^3}$ in volume. Therefore CdZnTe detectors are well suited for the development of compact and reliable radiation detection systems, and appealing for a large variety of applications, from nuclear medicine and radiation monitoring, to X-ray and gamma-ray astronomy.\\
The most significant drawback of CdZnTe devices, with respect to Si and Ge, is the comparatively poor transport properties of charge carriers, in particular holes. In order to operate the detectors with optimal performance, special electrode configurations which rely on the electron charge collection and are insensitive to the hole contribution, need to be implemented. Such configurations, also referred to as single charge carrier devices, include the virtual Frisch grid \cite{Cui_VirtualFrischGrid}, small pixel arrays or strips \cite{Shor_PixelStripCZT}, the co-planar grid \cite{Luke_Coplanar1,Luke_Coplanar2} and orthogonal strips \cite{McConnell_OrthogonalStrip} geometries. A complete review of these configurations and principle of operation can be found, for example, in \cite{He_CZTgeometries} or \cite{Zhang_CZTgeometries}.\\
In this work we characterize the performance of a pixelated CdZnTe detector custom designed for application on a Compton telescope, named \emph{MeVCube} \cite{Lucchetta_MeVCube}. The scientific payload is based on the CubeSat standard, a class of nanosatellites with precise restrictions and limitations in size and form factor \cite{CubeSat}. The \emph{MeVCube} instrument consists of $128$ CdZnTe detectors, arranged in a stack of two layers of $64$ detectors each; the separation of the two layers is $6\;\mathrm{cm}$. The telescope will scan the gamma-ray sky in the energy range from $200\;\mathrm{keV}$ to $4\;\mathrm{MeV}$, in a nearly equatorial low-Earth orbit, with
an altitude of $\sim 550\;\mathrm{km}$ and an inclination angle $\leq 5^{\circ}$. Soft gamma rays in this energy range interact through Compton scattering in the first layer of CdZnTe detectors, where the recoil electron is quickly absorbed. The scattered photon deposits the rest of the energy in the second layer\footnote{We do not consider Compton events in the same detector or detectors of the same layer, since these events leads to major uncertainties in the reconstruction of the direction of the incoming gamma ray.}. Measuring the energies of the scattered photon and of the recoil electron, together with their interaction positions, the energy and direction of the incoming gamma ray can be reconstructed. Further details about the \emph{MeVCube} geometry and its principle of operation are provided in a parallel paper \cite{Lucchetta_MeVCube}. Furthermore, the simulations performed in \cite{Lucchetta_MeVCube} have also shown that with an energy resolution from $\sim 6.5\%$ (FWHM) at $200\;\mathrm{keV}$ to $\lesssim 2.0\%$ above $1\;\mathrm{MeV}$, and a spatial resolution of $\mathcal{O}(2\,\mathrm{mm})$ in each direction for the interaction position, \emph{MeVCube} can reach an angular resolution of $\sim1.5^{\circ}$, a field-of-view of $\sim2\;\mathrm{sr}$ and a sensitivity level around $10^{-10}\;\mathrm{erg\, cm^{-2}\, s^{-1}}$.\\
Due to power constraints present in CubeSats, low-power read-out electronics for the detectors is mandatory. The VATA450.3 ASIC\footnote{Application Specific Integrated Circuit.}, developed by \emph{Ideas}\footnote{\emph{Integrated Detector Electronics AS}, \url{https://ideas.no/}.}, fulfils our requirements in terms of power consumption ($0.255\;\mathrm{mW/channel}$), dynamic range (up to $1.5$-$2\;\mathrm{MeV}$ deposited energy), noise and linearity (few percents). Further details about the read-out electronics are provided in Appendix~\ref{sec:VATA450.3}. Moreover, VATA450.3 has been used successfully for the read-out of CdTe sensors in the ASTRO-H space mission \cite{Tajima_VATA450}. We first present in more detail the experimental set-up employed for the measurements, and then present spectral and imaging performance of the detector.

\section{Pixel detector layout and experimental set-up}

\begin{figure}[htbp]
\centering
\includegraphics[width=0.70\textwidth]{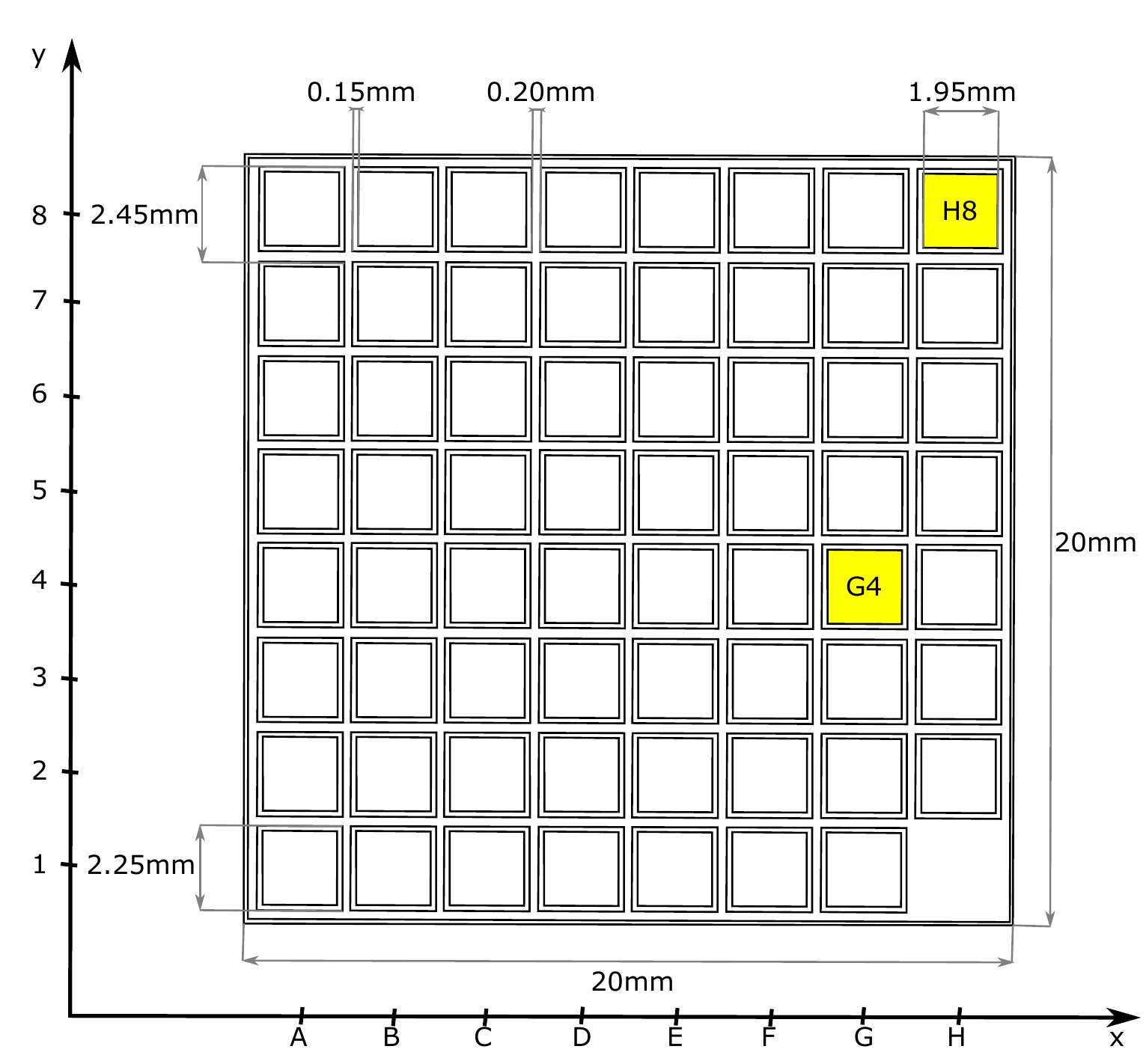}
\caption{\label{fig:AnodePattern} Schematic of the anode pattern of the CdZnTe detector as implemented in the \emph{MeVCube} project. Pixels are identified in a chess-like coordinate system, with rows ranging from $1$ to $8$, and columns ranging from A to H. The space of one pixel in the corner of the detector (H1) is left for the electrical contact of the steering grid. The pixels highlighted in yellow are considered in the following analysis to illustrate the performance of the detector. Pixel G4 is chosen as a representative of the best performing pixels, while H8 is selected as a representative of pixels with a comparatively poor performance, and is affected by distortions of the weighting potential and electric field typical for edge pixels (see Section~\ref{sec:DOIcorrection} for further details).}
\end{figure}

\begin{figure}[htbp]
\centering
\includegraphics[width=0.60\textwidth]{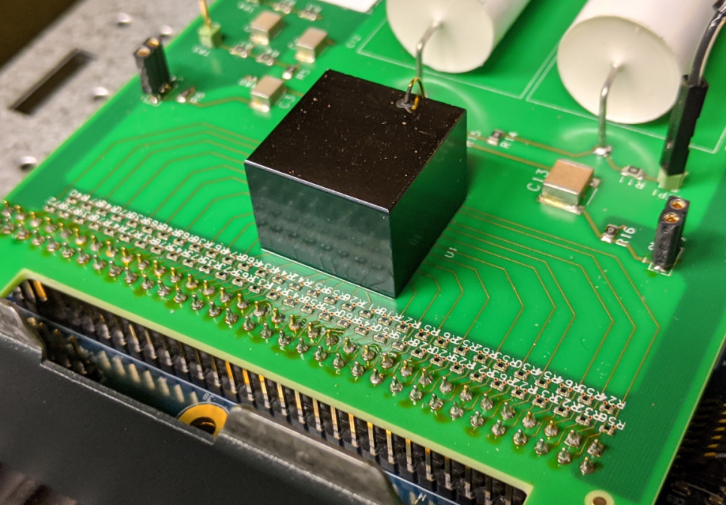}
\caption{\label{fig:Picture_pixelCZT} Close-up of the custom designed CdZnTe detector. The pixels are attached to a PCB; exposed on the top is the planar cathode. On the front row input connectors to the evaluation board provided by \emph{Ideas} are visible.}
\end{figure}

\noindent We characterize the performance of a pixelated CdZnTe detector, manufactured by \emph{Redlen Technologies}\footnote{\url{https://redlen.com/}.}, with a volume of $2.0\;\mathrm{cm} \times 2.0\;\mathrm{cm} \times 1.5\;\mathrm{cm}$ and a custom designed $8 \times 8$ pixel anode structure (see Figure~\ref{fig:AnodePattern}). The pixel size is $2.25\times 2.25\;\mathrm{mm^2}$ and the pixel pitch is $2.45\;\mathrm{mm}$. On the opposite side, the planar cathode is biased at $-2500\;\mathrm{V}$ (in our application). Gold contacts are used for both, the anode and the cathode.\\
\noindent According to the Shockley–Ramo theorem \cite{Shockley,Ramo} the signal induced in the electrodes can be computed based on quantities such as the weighting potential and the weighting field. In pixelated CdZnTe detectors, signals from the pixels are almost insensitive to the interaction position, while signals from the cathode linearly depend on the interaction depth. Therefore signals from triggered pixels provide a measurement of the energy deposited inside the detector, while the ratio between cathode and pixels signals provides a measurement for the depth-of-interaction. A detailed overview of signal formation in CdZnTe  and semiconductor detectors can be found in \cite{He_CZTgeometries}. A steering grid surrounds the pixels; the space of one pixel in the corner is used for the electrical contact of the steering grid. The presence of a steering electrode has shown to improve the charge collection efficiency of CdZnTe detectors, as reported, e.g., in \cite{SteeringGrid}. When the steering grid is biased to a slightly negative voltage with respect to the grounded pixels (around $-50\;\mathrm{V}$, see also Appendix~\ref{sec:SteeringGrid}), electrons are forced to move towards the pixels when approaching the anode surface, reducing charge loss in the pixels gap. The results of our measurements regarding the effect of the steering grid are reported in Appendix~\ref{sec:SteeringGrid}. The detector is attached using conductive epoxy to a $4$-layer PCB\footnote{Printed Circuit Board.}, hosting also filtering circuits for the high voltages of the cathode and the steering grid, and connectors to the VATA450.3 ASIC on its evaluation board for the pixels read-out (Figure~\ref{fig:Picture_pixelCZT}). A \emph{Galao} evaluation board configures the VATA450.3 ASIC for different working modes and controls data read-out and communication with the host computer. A comprehensive overview of the VATA450.3 ASIC, its principle of operation and experimental measurements verifying the manufacturer specifications are provided in Appendix~\ref{sec:VATA450.3}.

\begin{figure}[htbp]
\centering
\includegraphics[width=0.75\columnwidth]{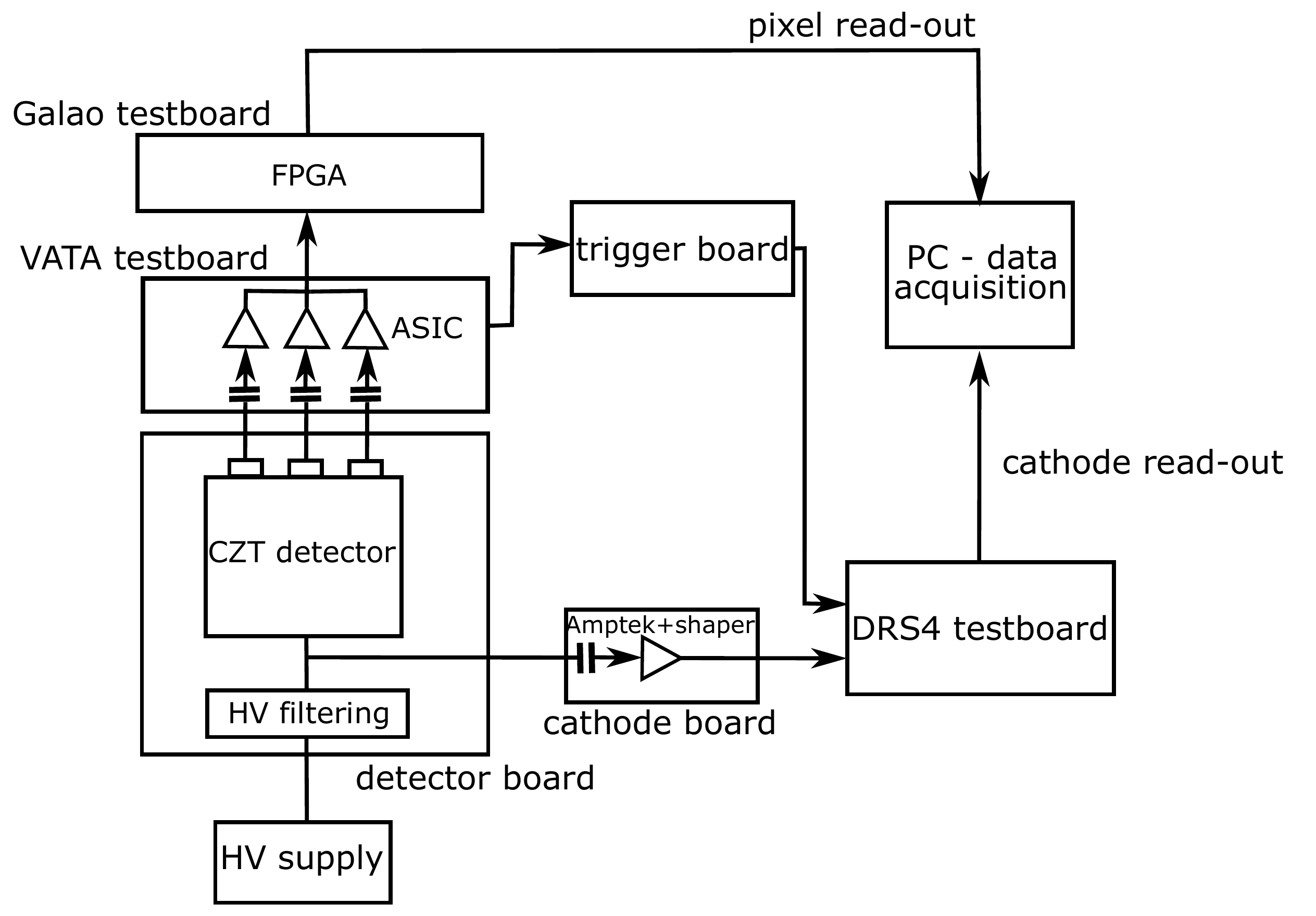}
\caption{\label{fig:BlockDiagram} Block diagram of the experimental set-up and read-out chain.} 
\end{figure}

\noindent A preliminary read-out system for the cathode is based on a discrete \emph{Amptek} A250F\footnote{\url{https://www.amptek.com/internal-products/a250f-and-a250fnf-high-density-charge-sensitive-preamplifiers}.} charge sensitive pre-amplifier, on a separate carrier board. Waveforms are sampled and recorded by a DRS4 (\emph{Domino Ring Sampler}, \cite{DRS4}) ASIC on its evaluation test-board\footnote{\url{https://www.psi.ch/en/drs/evaluation-board}.}, for signal and baseline analysis. At a later stage of the project the entire read-out will be based on VATA450.3 ASICs\footnote{A read-out system completely based on VATA450.3 would require the implementation of two ASICs, the development of dedicated carrier boards and our own first version of read-out firmware. The work is beyond the scope of this paper and will be carried out in the next stages of the \emph{MeVCube} project.}. Only information on the deposited charge is used in the following analysis; no additional information or corrections involving cathode drift time or cathode-to-anode time difference has been implemented in the current set-up, in order to develop a framework consistent with a read-out system completely based on the VATA450.3. The pixels and the cathode are AC coupled to the read-out electronics. A block diagram of the experimental set-up and read-out chain is shown in Figure~\ref{fig:BlockDiagram}.\\
The high resistivity, low bulk-capacitance and surface passivation of the CdZnTe and contacts result in a low dark current. However a large parasitic capacitance is expected in the current setup, due to the size of the evaluation boards and long traces in the PCB. Details about noise measurements for our setup are given in Appendix~\ref{sec:VATA450.3}. An optimization of the layout of the carrier boards, in order to improve the noise performance of the set-up is foreseen for the future.\\
The performance of the detector is evaluated using radioactive gamma-ray sources. Different radioactive sources were used, in order to test the spectral response of the detector on a wide energy range: a Cs-$137$ source with a gamma energy line of $662\;\mathrm{keV}$, a Ba-$133$ source with lines at $81$, $276.4$, $302.9$ and $356\;\mathrm{keV}$, and a Co-$60$ source with lines at $1.17$ and $1.33\;\mathrm{MeV}$. The detector has been tested at room temperature using a bias voltage of $-2500\;\mathrm{V}$.

\section{Depth-of-interaction correction}
\label{sec:DOIcorrection}
The first radioactive source used for detector characterization is a Cs-$137$ source. In order to characterize the performance of each pixel individually, we analyze events which trigger a single anode pixels (in coincidence with the cathode signal). Events in which two or more pixels are triggered simultaneously, are rejected from our analysis. For the Cs-$137$ source, such events constitute about $15\%$ of the data sample. 

\begin{figure}[ht]
	\centering
	\subfloat[Pixel G4.]{
	\includegraphics[width=0.49\textwidth]{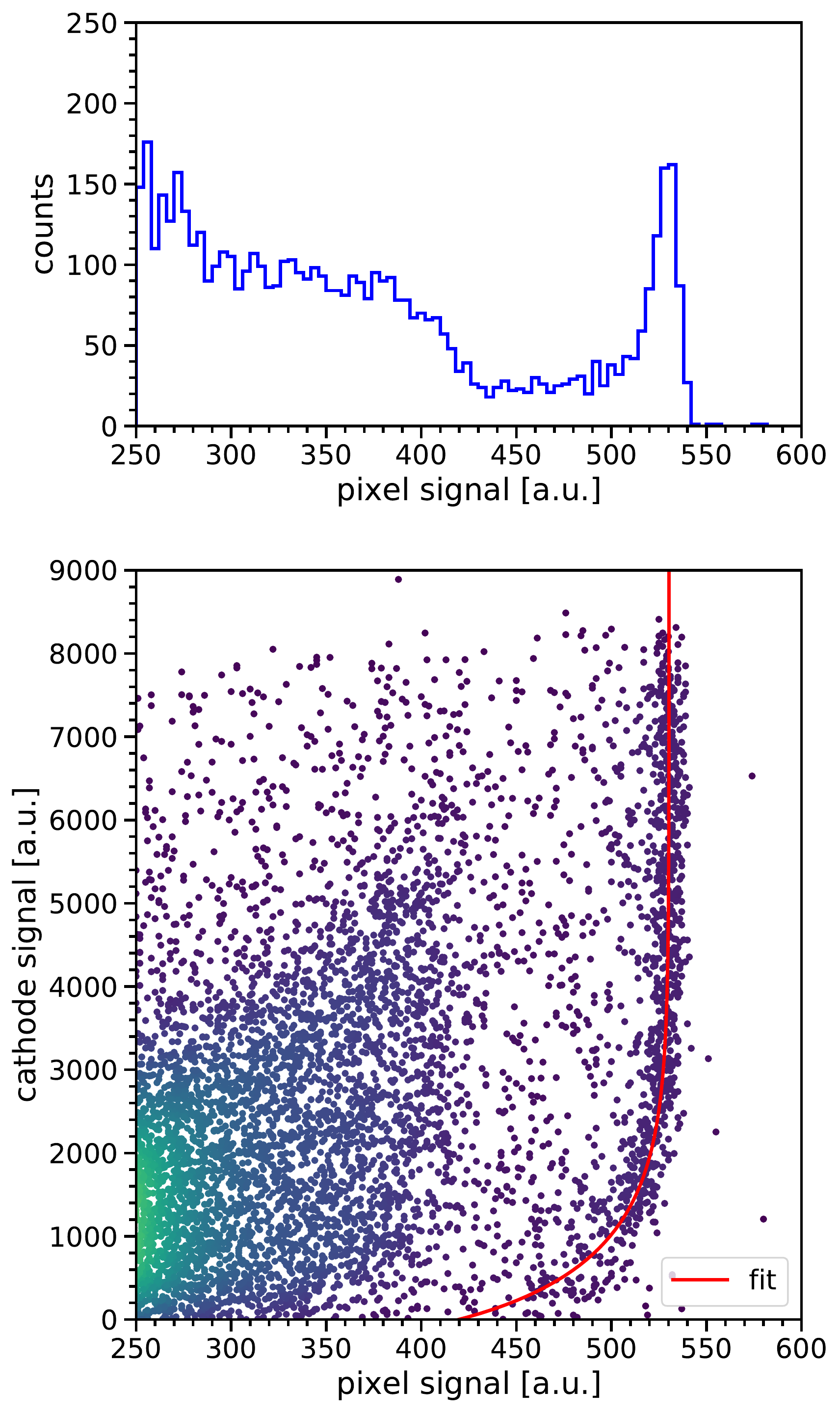}}
	\vspace{\floatsep}
	\subfloat[Pixel H8.]{
	\includegraphics[width=0.49\textwidth]{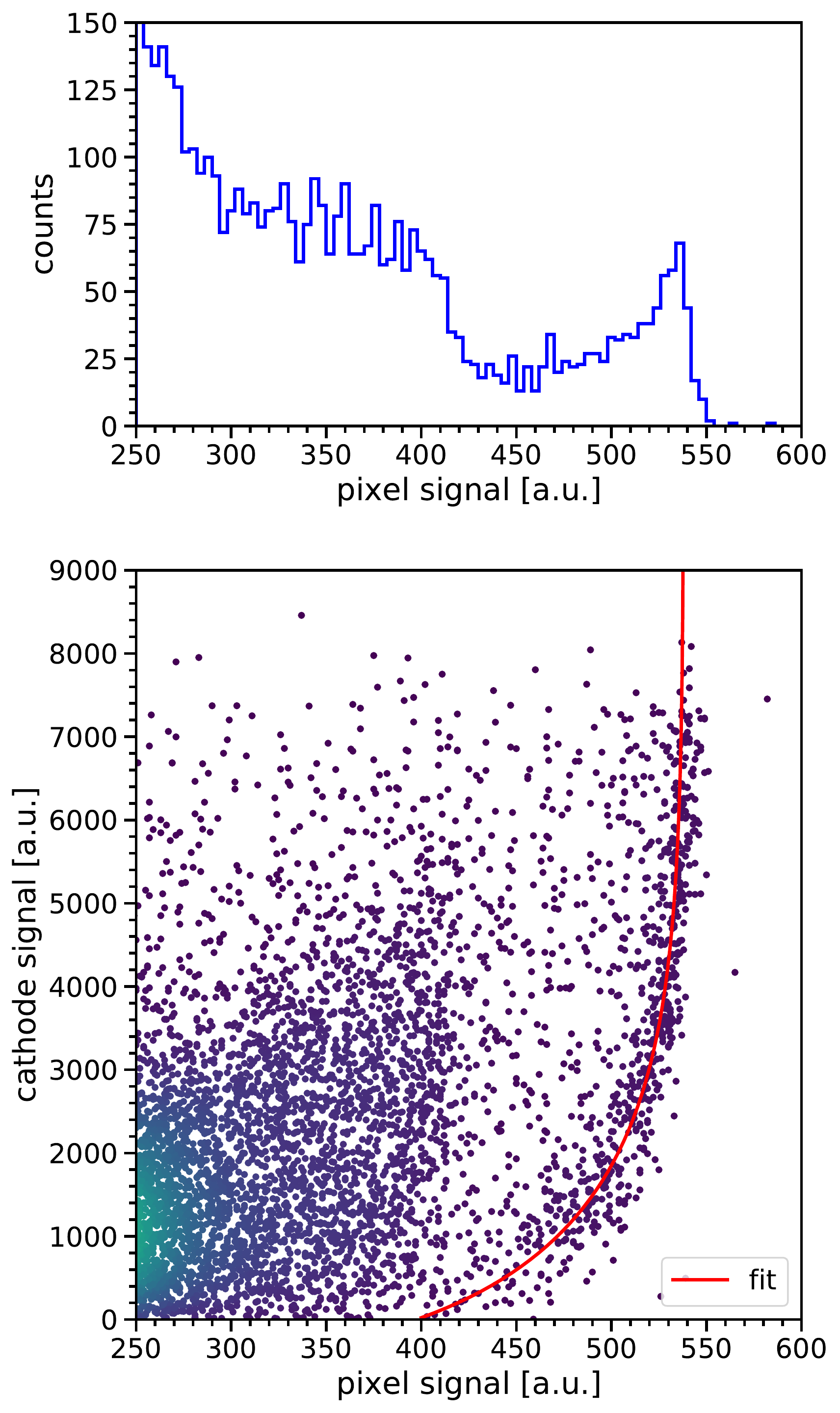}}
	\caption{\label{fig:DoiCorrection1} Top: energy spectra of the Cs-$137$ radioactive source for two selected pixels, showing very different behaviour. Bottom: enhanced spectral performance can be achieved through a depth-of-interaction technique: signals from the cathode, taken in coincidence with those of the pixels, provide a correction for incomplete charge collection and trapping effects.}
\end{figure}

\begin{figure}[ht]
	\centering
	\subfloat[Pixel G4.]{
	\includegraphics[width=0.49\textwidth]{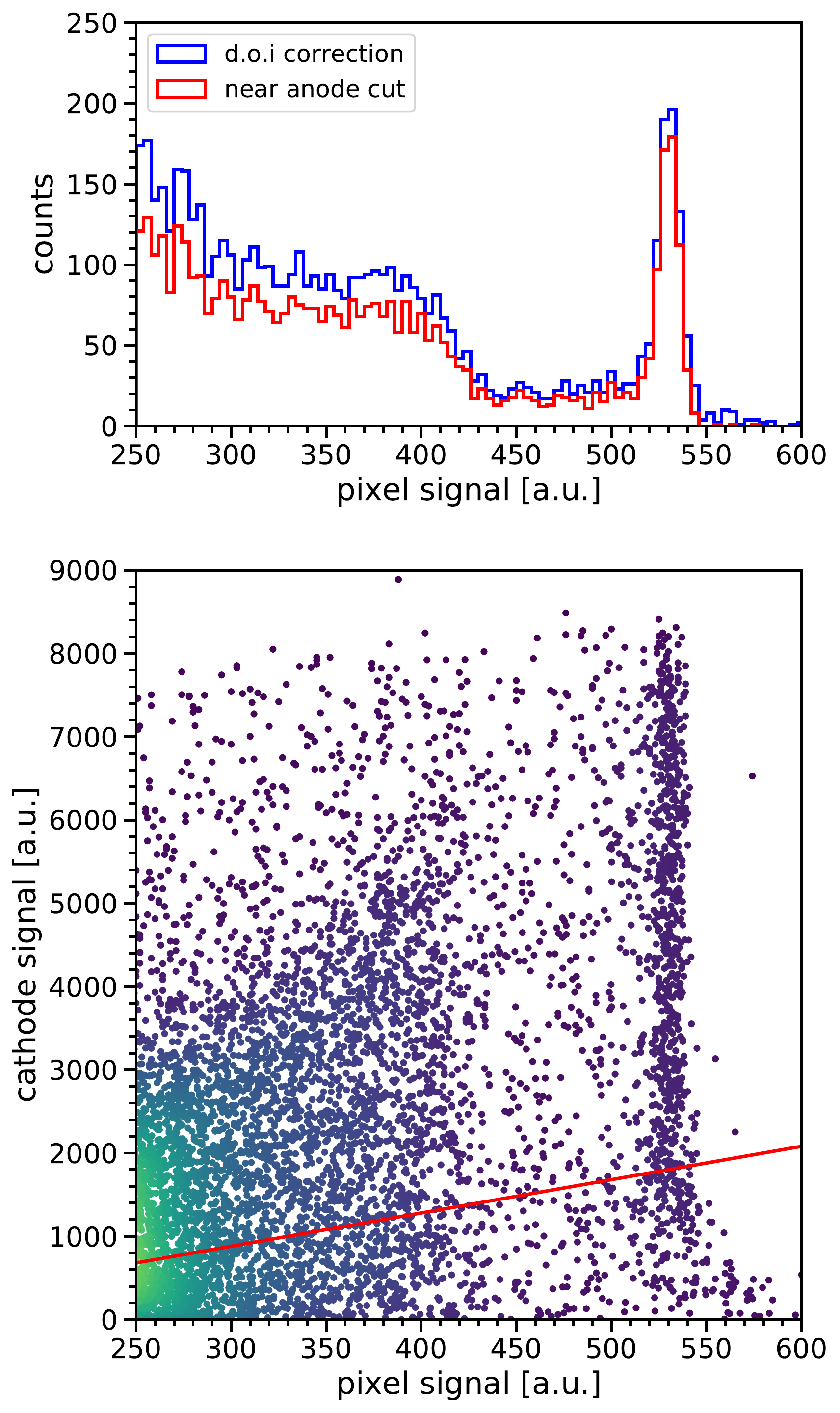}}
	\vspace{\floatsep}
	\subfloat[Pixel H8.]{
	\includegraphics[width=0.49\textwidth]{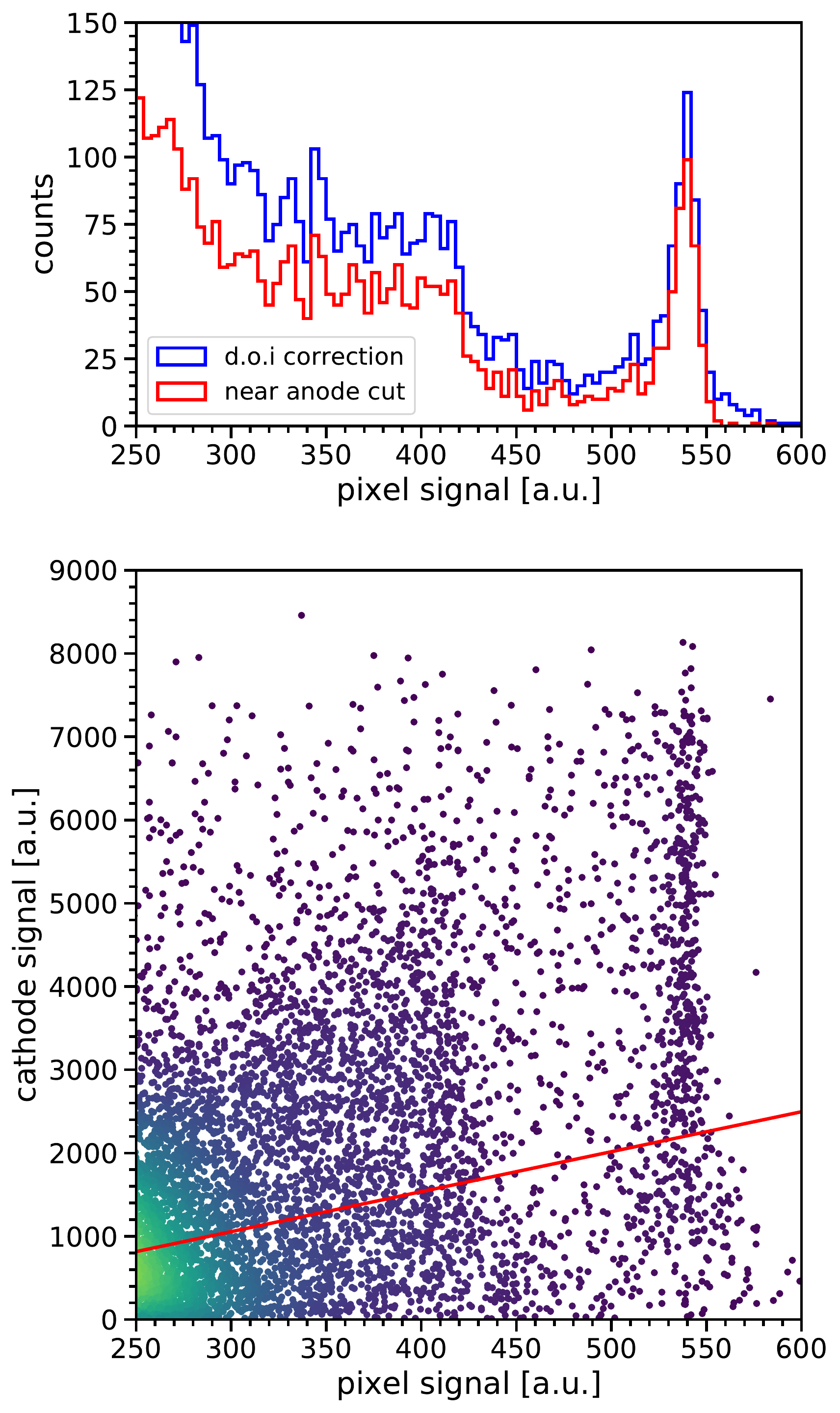}}
	\caption{\label{fig:DoiCorrection2} Bottom: scatter plot between the cathode signal and the new pixel signal obtained after depth-of-interaction correction. Top: the linearisation of the correlation profile leads to a sharper energy line for the Cs-$137$ photo-peak. A further improvement of the spectra can be achieved by imposing a cut in the interaction position, marked by the red lines in the scatter plots. The cut is introduced in order to reject events close to the anode surface, affected by the greatest distortions, with uniform efficiency in energy.}
\end{figure}

\noindent The spectral results for two selected pixels are shown in Figure~\ref{fig:DoiCorrection1}. A common behaviour observed for the measured spectra of thick CdZnTe detectors is the presence of pronounced tails towards lower energies with respect to the photo-peaks (see, for example, \cite{Yang_PixelCZT} or \cite{Shor_PixelStripCZT}). The effect is due to incomplete charge collection and charge carrier trapping; imbalances of the electric field and potential for pixels on the edge of the detector might also contribute. Since the low-energy tails are a consequence of depth-dependent charge collection efficiency, depth information can be used to correct the signals, enhancing the overall spectral performance of the detector. This is achieved by the read-out of signals from the planar cathode, taken in coincidence with those of the anode pixels. The bottom of Figure~\ref{fig:DoiCorrection1} shows the measured relationship between the two signals. The distinct correlation profile of the $662\;\mathrm{keV}$ photo-peak line can be linearised in order to enhance the overall spectral performance of the detector, in a procedure called \emph{depth-of-interaction correction} \cite{Pamelen_CZTcorrection,Shor_PixelStripCZT}. A proper correction is provided by the following function:
\begin{equation}
E = a \cdot \frac{S_{\mathrm{pixel}}}{1- \exp{\bigg( - b\cdot \frac{S_{\mathrm{cathode}}}{S_{\mathrm{pixel}}} \bigg)}} \, ,
\label{eq:doi_correction}
\end{equation}
where $S_{\mathrm{cathode}}$ and $S_{\mathrm{pixel}}$ represent the integrated charge measured at the cathode and the pixel, respectively, and $a$ and $b$ are the fitting parameters.\\
The relationship between the cathode signal and the corrected pixel signal, obtained after depth-of-interaction correction is highlighted in the scatter plot of Figure~\ref{fig:DoiCorrection2}. As observed from the blue histograms in Figure~\ref{fig:DoiCorrection2}, the reduction of the low-energy tails is remarkable and the photo-peak line is more pronounced. The peak-to-valley ratio improves from $5.9$ to $8.2$ for the G4 pixel, and from $2.8$ to $6.6$ for the H8 pixel.

\noindent A degradation of the energy response of the detector is still observed for interactions very close to the anode surface (smaller than the pixel size), corresponding to the smallest values of the cathode signal. Since the fitting function in \eqref{eq:doi_correction} only takes into account the trapping effect in the case of an ideal weighting potential, it does not provide an accurate description for interaction closer to the anode side, which might be affected by imbalances in potentials and electric field. Therefore the spectral resolution shown can be improved further, selecting interactions with:
\begin{equation}
\frac{S_{\mathrm{cathode}}}{S_{\mathrm{pixel}}} > \mathrm{threshold} \, .
\label{eq:cut_PixelAnalysis}
\end{equation}

\begin{figure}[ht]
	\centering
	\includegraphics[width=0.65\textwidth]{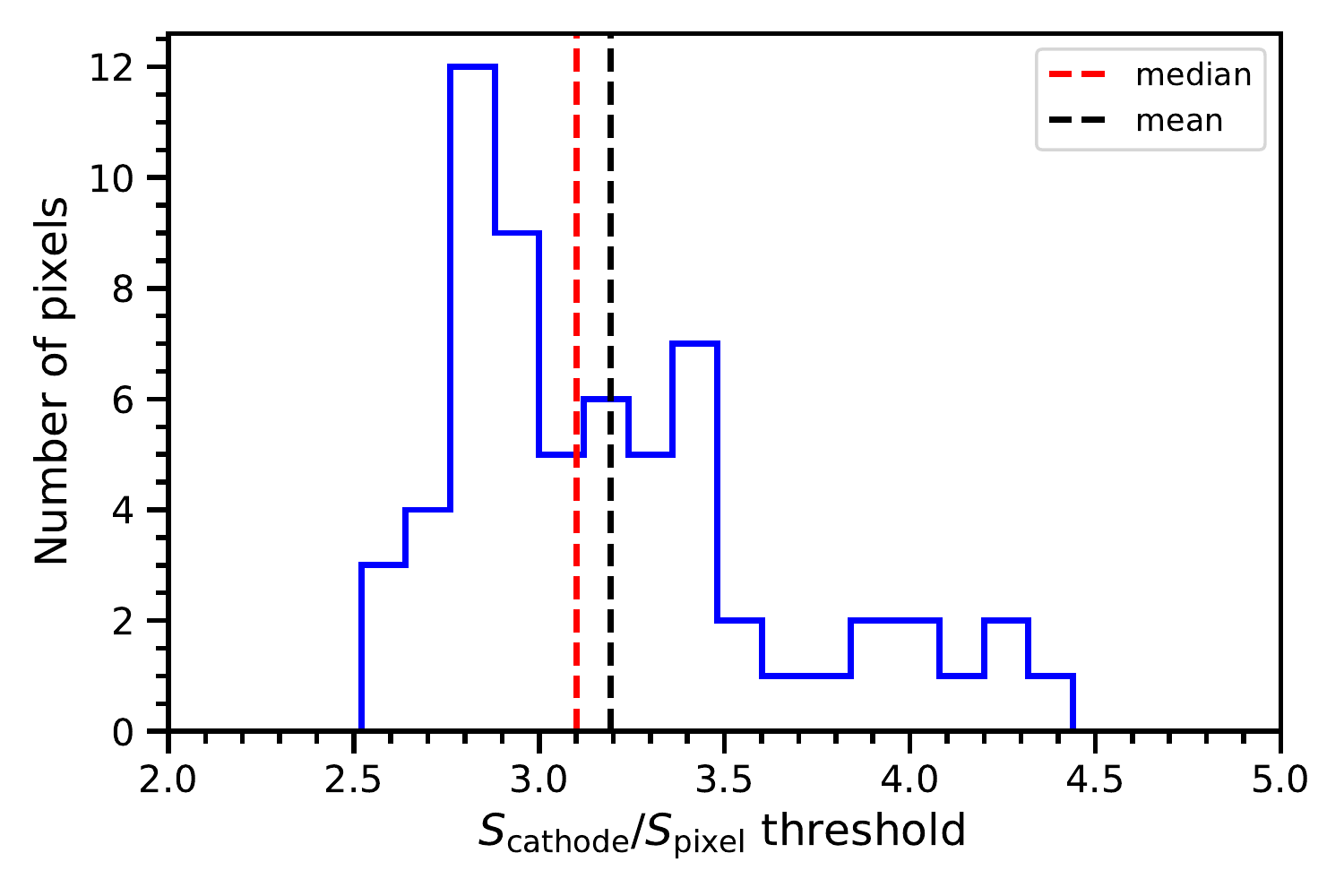}
	\caption{\label{fig:ThresholdDistribution} Distribution of the $S_{\mathrm{cathode}}/S_{\mathrm{pixel}}$ threshold value for all pixels; events below the threshold are removed from the analysis. The median and the mean of the distribution are displayed with the red and black vertical lines, respectively.}
\end{figure}

\noindent The value of the threshold is tuned channel by channel, in order to take into account the different distortions and behaviour of each pixel. The distribution of the threshold value for all pixels is provided in Figure~\ref{fig:ThresholdDistribution}. The highest values of the threshold are found for pixels at the edges and corners of the detector, affected by the largest distortions in the weighting potential and electric field. Since the ratio between the cathode and pixel signals is equal to the interaction depth (scaled to the total detector thickness), the relation is a cut in the interaction position, meant to provide uniform efficiency over the energy range of interest. The final spectra obtained for the two selected pixels are shown by the red histograms in Figure~\ref{fig:DoiCorrection2}. The imposed cut limits the active volume of the detector to $\sim 80\%$, taking also into account that the space of one pixel is left for the electrical contact of the steering grid. In terms of efficiency, $13\%$ of the counts are lost for pixel G4, while $22\%$ of the counts are lost for the pixel H8.

\section{Energy resolution of the CdZnTe detector}
\begin{figure}[htbp]
	\centering
	\includegraphics[width=0.70\textwidth]{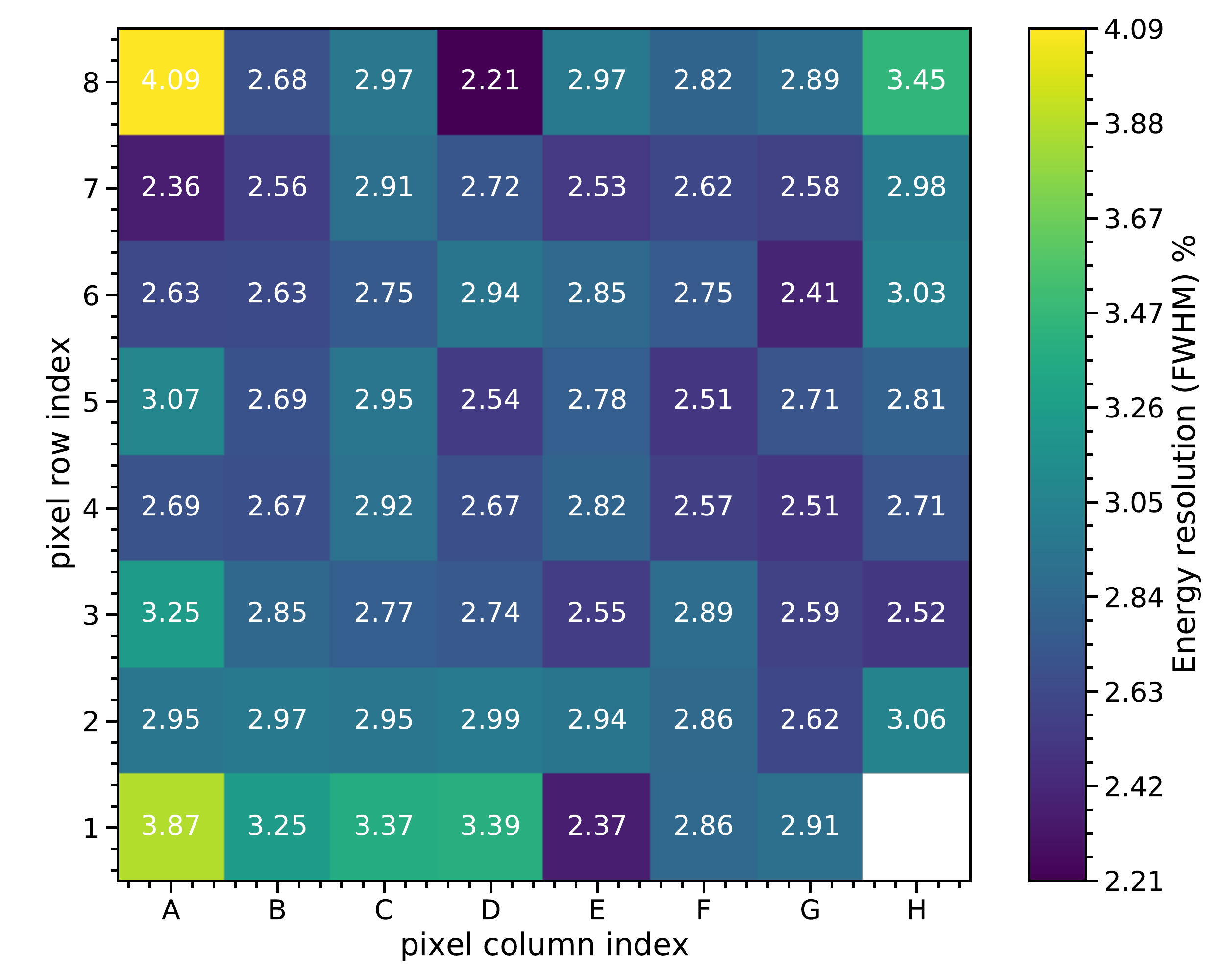}
	\caption{\label{fig:EnRes_Cesium} Energy resolution (expressed in FWHM) measured for the $662\;\mathrm{keV}$ line of Cs-$137$.}
\end{figure}

Figure~\ref{fig:EnRes_Cesium} displays the energy resolution measured at $662\;\mathrm{keV}$ for all pixels, expressed in \emph{full width at half maximum} (FWHM). As it can be seen, the spectral performance is very uniform throughout the detector: the energy resolution measured for all pixels in the bulk of the detector is $\lesssim 3.0\%$ at $662\;\mathrm{keV}$, while few pixels in the edges and corners of the detectors exhibit an energy resolution $>3.0\%$. The deterioration of the energy resolution for edge pixels is expected due to distortions of the electric field at the boundaries of the detector.

\begin{figure}[htbp]
	\centering
	\includegraphics[width=0.65\textwidth]{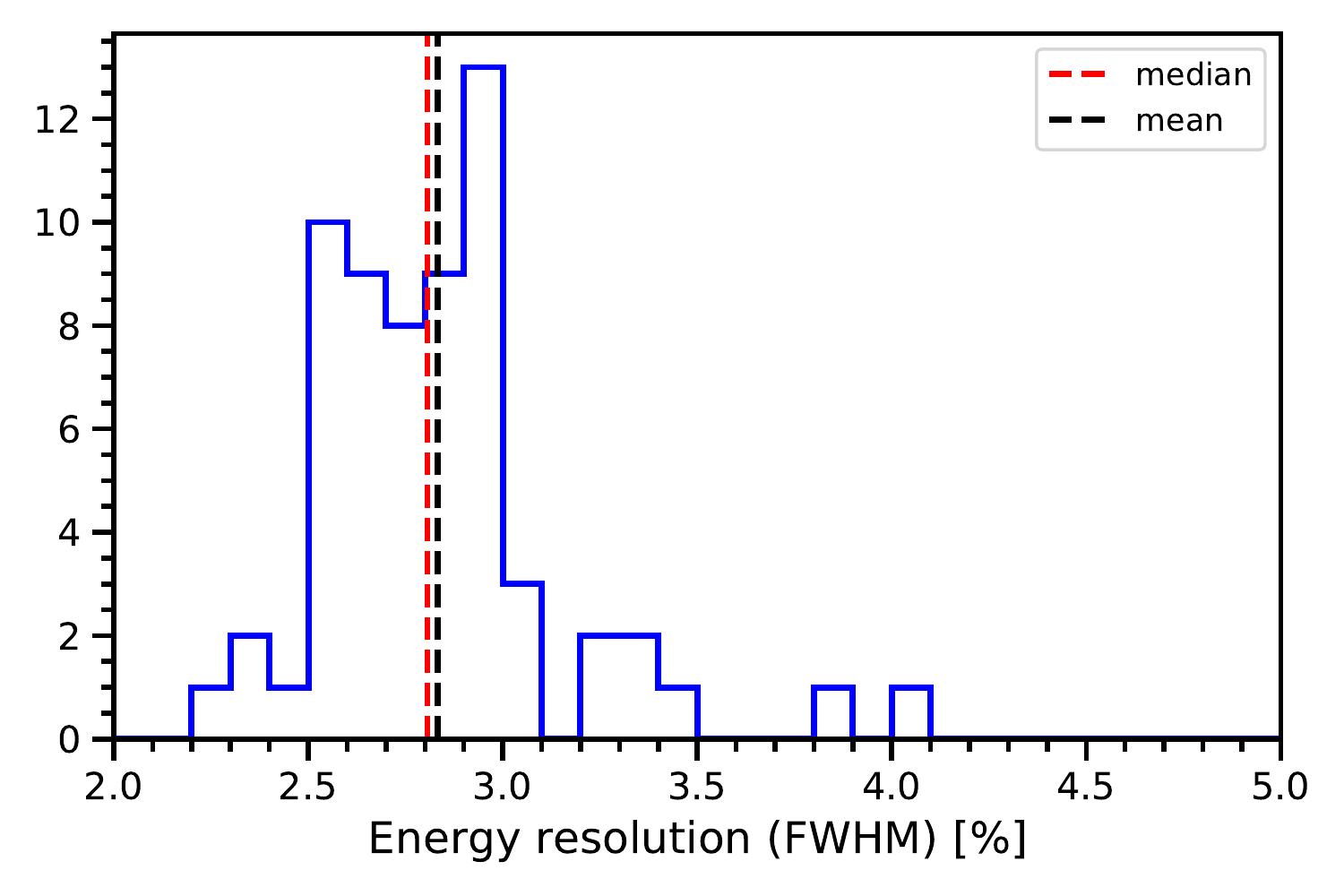}
	\caption{\label{fig:CsDistributionMedian} Distribution of the measured energy resolution at $662\;\mathrm{keV}$ for all pixels. The median and the mean of the distribution are displayed with the red and black vertical lines, respectively.}
\end{figure}

\noindent An overall ``global'' energy resolution for the detector is provided by the median over all $63$ pixels. Figure~\ref{fig:CsDistributionMedian} shows the distribution of the measured energy resolution in all pixels, with the median and mean marked respectively by the red and black vertical lines; the median value is $\sim 2.8\%$.

\begin{figure}[htbp]
	\centering
	\includegraphics[width=0.65\textwidth]{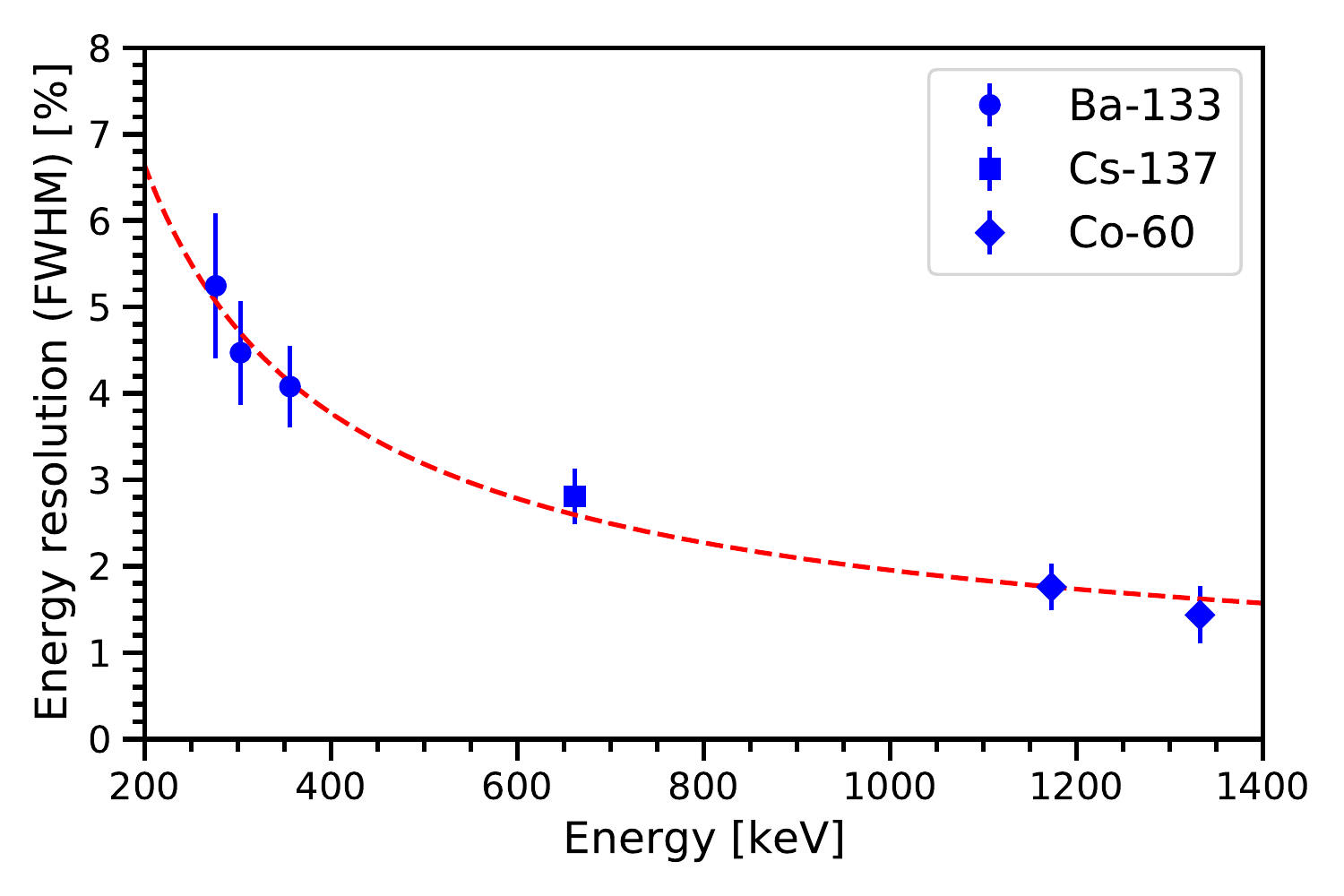}
	\caption{\label{fig:Pixel_EnRes} Energy resolution of the pixelated CdZnTe detector as a function of energy, measured for different radioactive sources. For each gamma-ray line the resolution is obtained as the median value of the distribution for all $63$ pixels.}
\end{figure}

\noindent In order to characterize the detector performance over a wider energy range, we made use of other radioactive sources, such as Ba-$133$ and Co-$60$, allowing for the measurement of the energy resolution from $\sim 200\;\mathrm{keV}$ to above $1\;\mathrm{MeV}$. Specifically the detectable energy lines are: $276.4\;\mathrm{keV}$, $302.9\;\mathrm{keV}$ and $356\;\mathrm{keV}$ for Ba-$133$ and $1.17\;\mathrm{MeV}$ and $1.33\;\mathrm{MeV}$ for Co-$60$. Measurements of the $81\;\mathrm{keV}$ line from Ba-$133$ is only possible for the pixels at the sides of the detector directly facing the radioactive source, since at this energy $99.9\%$ of the total radiation is stopped after $\sim 0.4\;\mathrm{cm}$ in CdZnTe\footnote{The attenuation coefficient for CdZnTe can be computed at \url{https://physics.nist.gov/PhysRefData/Xcom/html/xcom1.html}}. For these pixels an energy resolution between $15$ and $20\;\mathrm{keV}$ FWHM is measured, mainly limited by electronic noise. In fact the total noise contribution from the measurement of the width of the pedestal distribution is $15.4\;\mathrm{keV}$ FWHM, averaged for all pixels (see Figure~\ref{fig:NoiseFinalSetUp} in Appendix~\ref{sec:VATA450.3}). The spectral performance of the detector, as a function of the energy of the gamma photo-peak lines, is plotted in Figure~\ref{fig:Pixel_EnRes}: the measured energy resolution is on average $\sim6.5\%$ around $200\;\mathrm{keV}$ decreasing to $\lesssim2\%$ at energies above $1\;\mathrm{MeV}$.

\section{Depth resolution of the pixelated CdZnTe detector}
\begin{figure}[ht]
	\centering
	\includegraphics[width=0.70\textwidth]{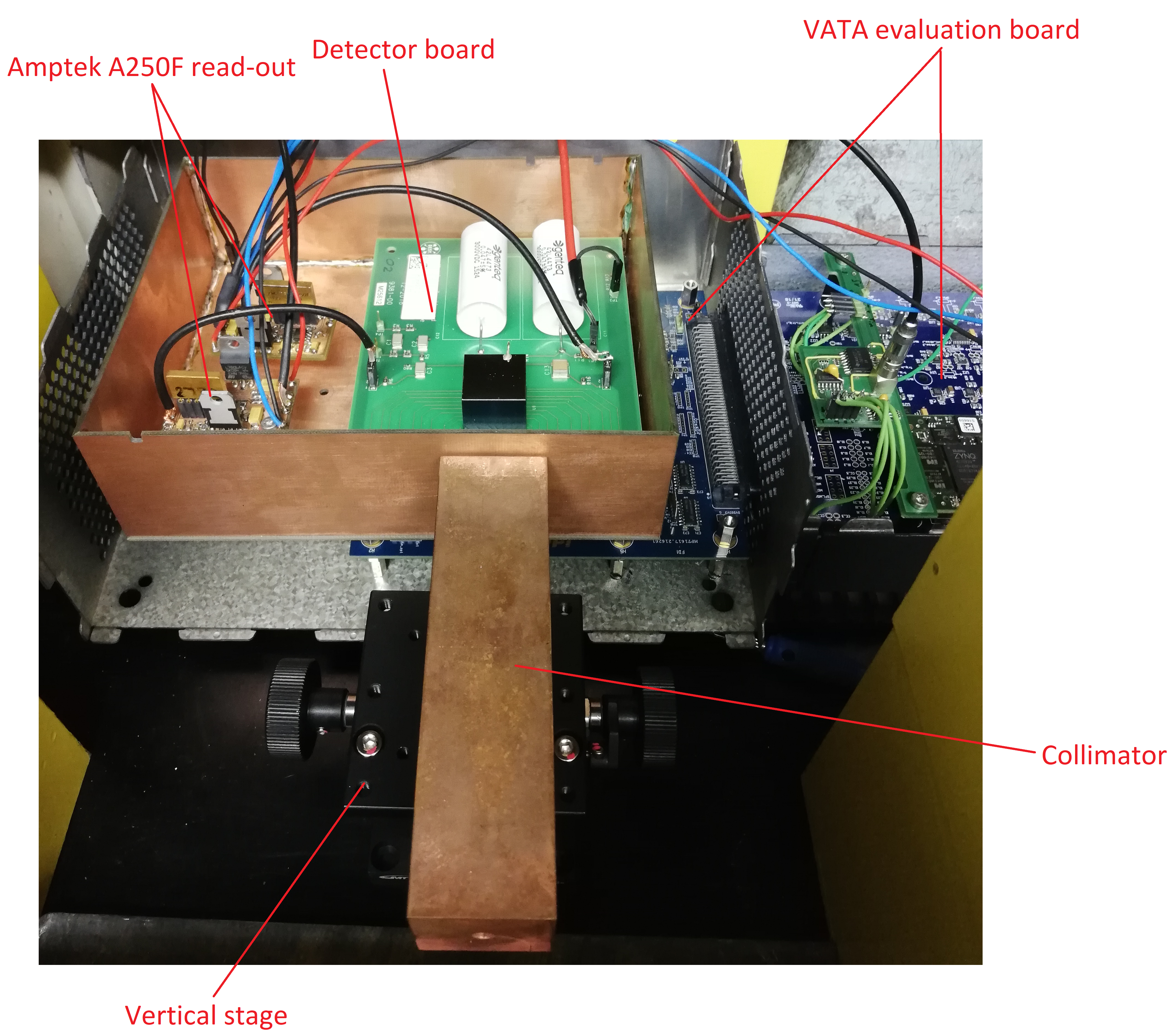}
	\caption{\label{fig:PictureCollSetUp} Overview of the experimental set-up employed for the depth of interaction measurements.}
\end{figure}

A second requirement for our application is a $\mathcal{O}(2\,\mathrm{mm})$ $3$-D spatial resolution in each direction. The spatial resolution on the anode plane (interaction on the x-y plane) is dictated by the pixel pitch, equal to $2.45\;\mathrm{mm}$, while the interaction depth (interaction location on the z axis) can be reconstructed from the ratio between the cathode and the pixel signals. The depth resolution of the detector is investigated with a Cs-$137$ radioactive source and a copper collimator, manufactured at DESY (see Figure~\ref{fig:PictureCollSetUp}).

\begin{figure}[htbp]
	\centering
	\includegraphics[width=0.75\textwidth]{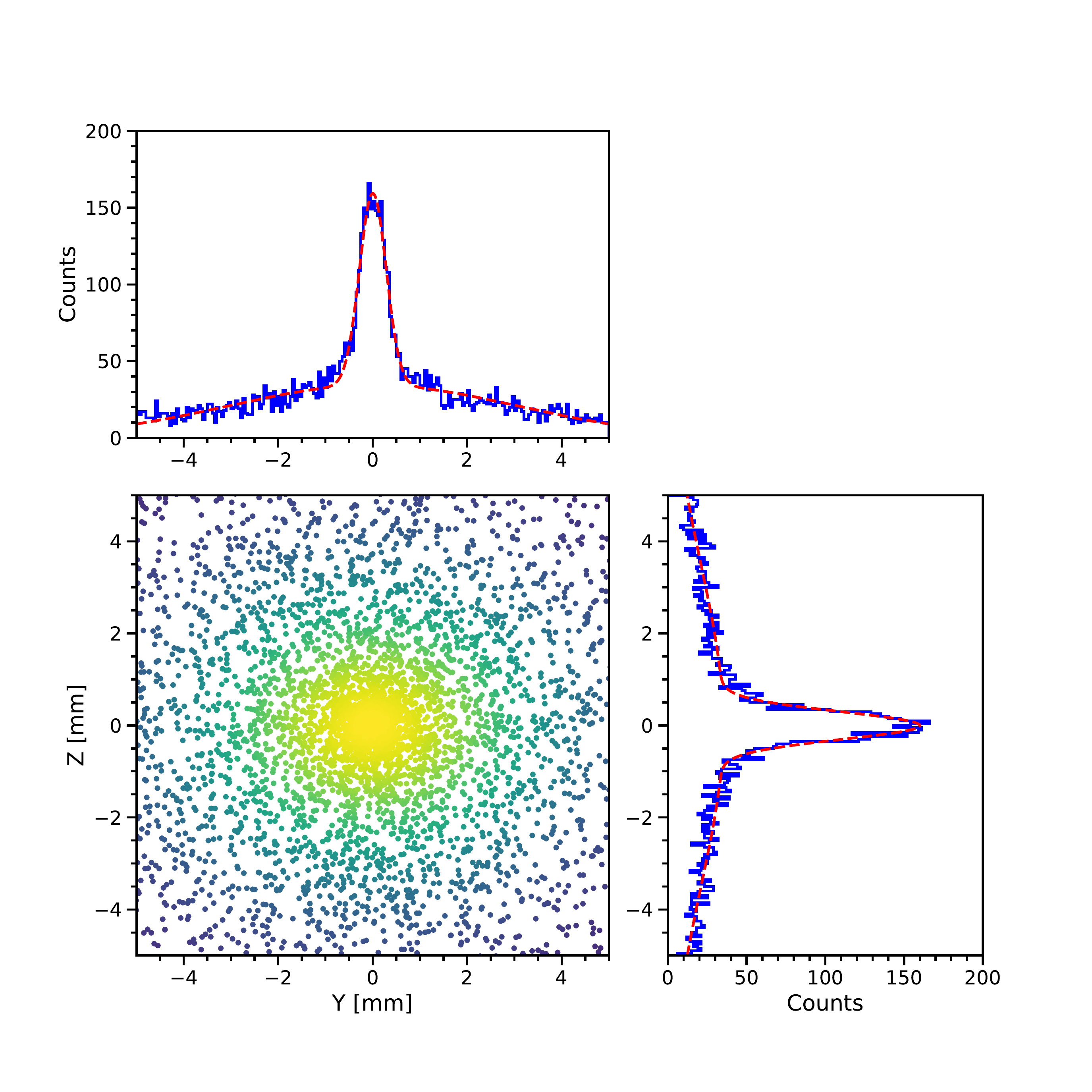}
	\caption{\label{fig:CollimatorSim} Interaction position as obtained from the \emph{Geant4} simulation with the collimator set-up, considering an ideal CdZnTe detector with infinitely precise energy and spatial resolution.}
\end{figure}

\begin{figure}[htbp]
	\centering
	\includegraphics[width=0.65\textwidth]{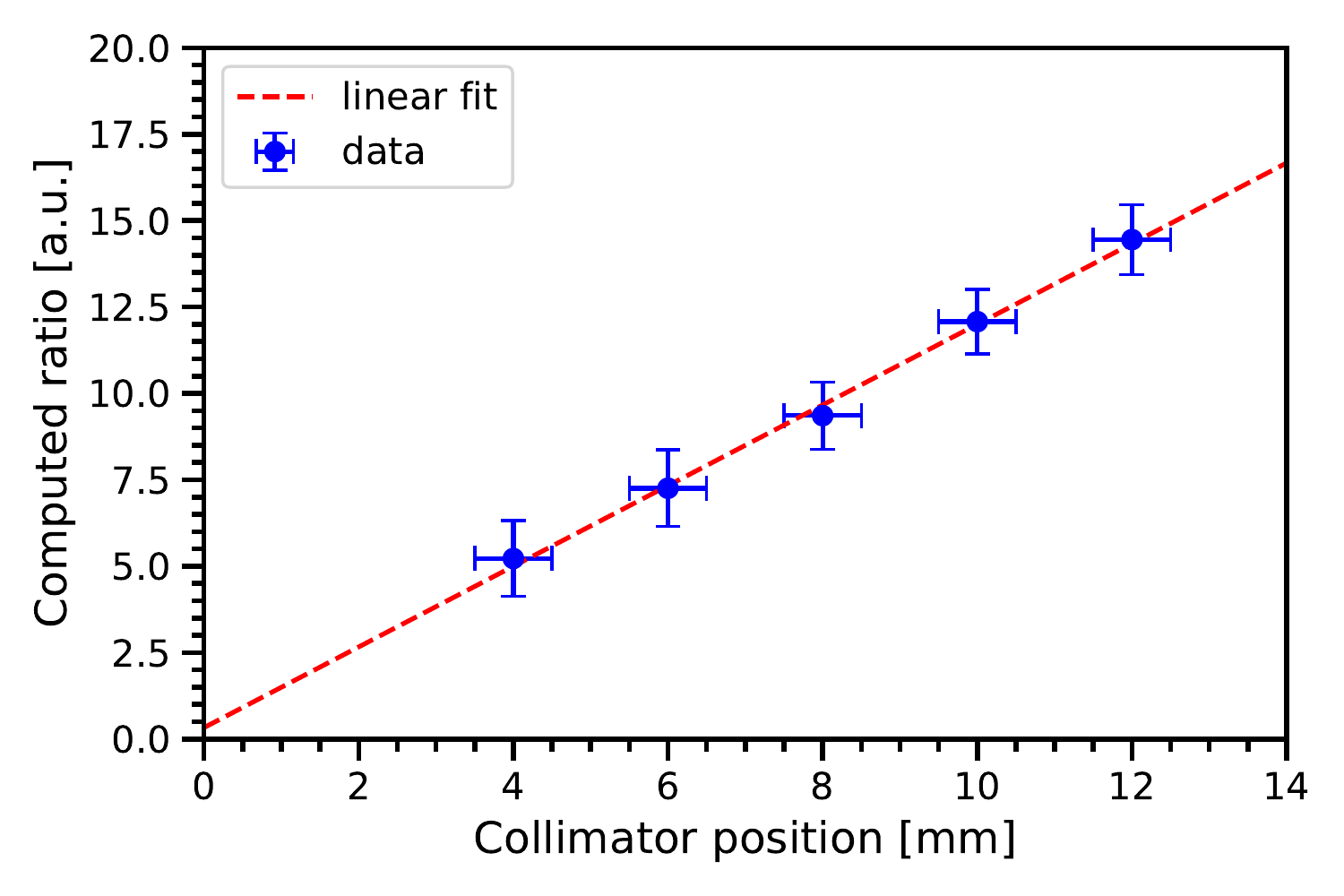}
	\caption{\label{fig:CollimatorLinearFit} Linear relation between the collimator position and the computed ratio between the cathode and pixel signals. Errors on the x axis are given by the scale of the vertical stage, while the widths of the reconstructed depth distributions are provided for the y axis.}
\end{figure}

\noindent The collimator has a length of $10\;\mathrm{cm}$ and a drill hole of $0.5\;\mathrm{mm}$ in diameter. In the current set-up the collimator stands to a distance of $\sim3\;\mathrm{cm}$ from the detector. A \emph{Geant4} \cite{Agostinelli_Geant4} simulation has been performed, in order to evaluate the geometrical effects of the set-up on the spatial resolution, due to the finite size of the collimator beam and the distance between the collimator and the detector. For this purpose, an ideal detector with infinitely precise energy and spatial resolution has been considered. The results of the simulations are shown in Figure~\ref{fig:CollimatorSim}.\\
Pronounced tails superimposed to the gamma-ray peak can be observed, due to inefficient collimator shielding in a penumbra region around the hole\footnote{In, for narrow angles around the collimator's hole a non-negligible fraction of the passing gammas can pass the copper material around the exit of the hole.}. A proper fit of the data in Figure~\ref{fig:CollimatorSim} is achieved with a double gaussian distribution, with identical mean value: the first one accounts for the ``background'' in the penumbra region, while the second one gives the desired collimator spatial resolution, $R_{\mathrm{c}}$. The measured depth-of-interaction resolution of the system is given by:
\begin{equation}
R_{\mathrm{sys}} = \sqrt{R^2_{\mathrm{det}} + R^2_{\mathrm{c}}}\, ,
\label{eq:DepthResFormula}
\end{equation}
where $R_{\mathrm{det}}$ is the intrinsic spatial resolution of the detector and $R_{\mathrm{c}}$ the component due to the collimator geometry, evaluated through simulation.\\
The depth resolution of the CdZnTe detector has been measured for different scanning positions, adjusted via a vertical stage with a stroke of $20\;\mathrm{mm}$ and a scale of $1\;\mathrm{mm}$. The interaction depth is provided by the ratio between the cathode and the anode signals (corrected with the depth-of-interaction technique) as in \eqref{eq:cut_PixelAnalysis}, from those events corresponding to the Cs-$137$ photo-peak. The relation between the computed ratio and the collimator position is provided by Figure~\ref{fig:CollimatorLinearFit}.

\begin{figure}[htbp]
	\centering
	\includegraphics[width=0.65\textwidth]{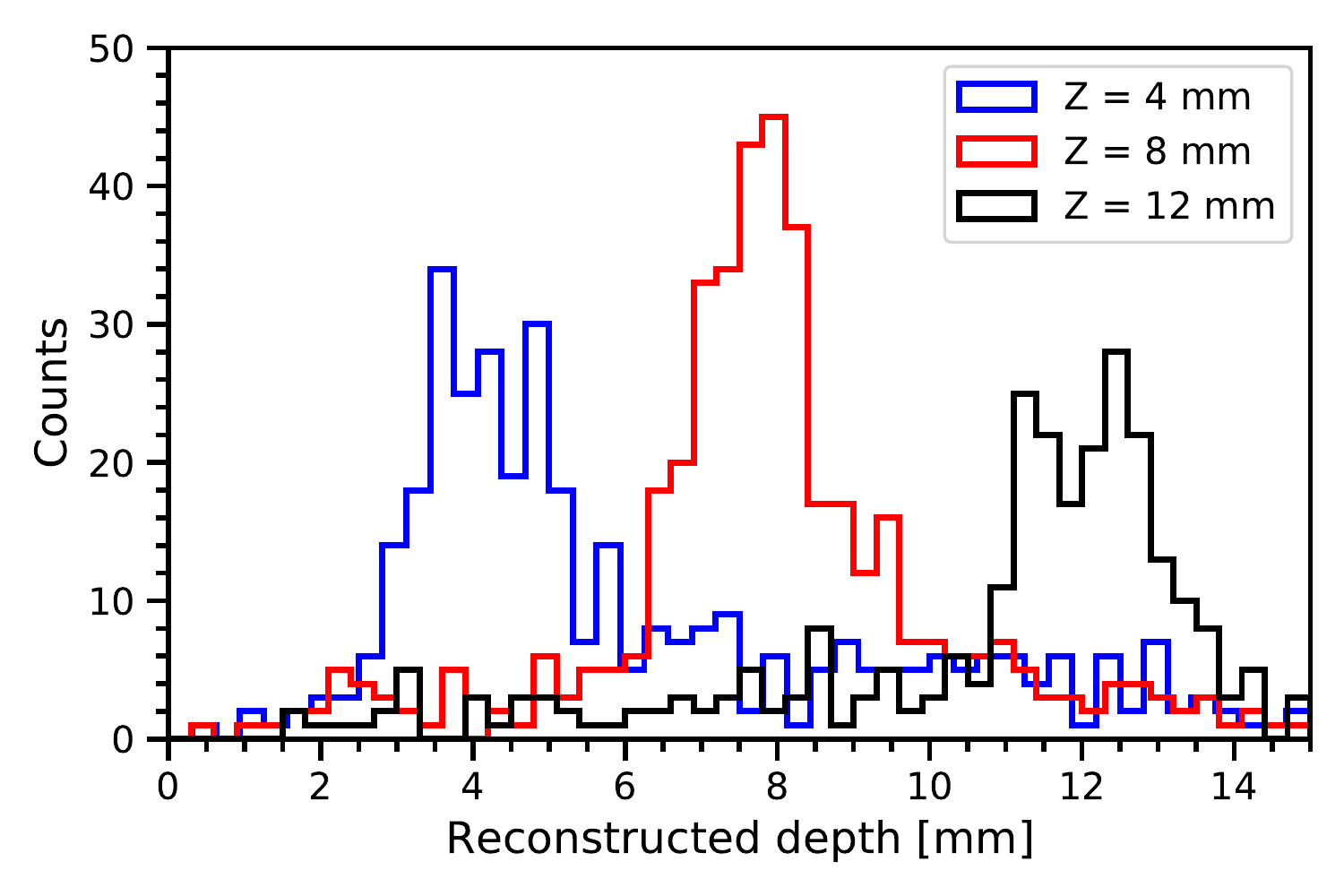}
	\caption{\label{fig:ReconstructedDepth} Reconstructed interaction depth for three different scanning positions.}
\end{figure}

\noindent An example of the reconstructed depth for three different scanning positions is shown in Figure~\ref{fig:ReconstructedDepth}. After subtraction of the geometrical component evaluated through simulations, described in 
\eqref{eq:DepthResFormula}, an average depth resolution of $\sim1.8\;\mathrm{mm}$ (in FWHM) is achieved for the detector, meeting the requirements of the \emph{MeVCube} project.

\section{Conclusions}
The performance of a $2.0\;\mathrm{cm} \times 2.0\;\mathrm{cm} \times 1.5\;\mathrm{cm}$ pixelated CdZnTe detector has been evaluated. In our current set-up pixels are read-out by the low-power ASIC VATA450.3, while a read-out system based on the \emph{Amptek} A250F charge sensitive pre-amplifier has been implemented for the cathode. Experimental measurements have shown that this combination can cover the energy range between $200\;\mathrm{keV}$ and $2.0\;\mathrm{MeV}$, with an acceptable noise and an integral non-linearity of just a few percent.\\
Energy and spatial resolution of the CdZnTe detector has been characterized by irradiation with different radioactive sources. A depth of interaction correction has been implemented, in order to obtain optimal charge collection and improved performance. After depth-of-interaction correction, an energy resolution of $\lesssim 3.0\%$ in FWHM is achieved at $662\;\mathrm{keV}$, with a median value of $2.8\%$; $10$ pixels, located at the edges of the detector exhibit an energy resolution $>3.0\%$. The median energy resolution decreases to about $\sim6.5\%$ at $200\;\mathrm{keV}$ and increases to $\lesssim 2.0\%$ at energies above $1\;\mathrm{MeV}$. The spatial resolution on the anode plane is dictated by the pixel pitch ($2.45\;\mathrm{mm}$), while a depth resolution of $\sim 1.8\;\mathrm{mm}$ (FWHM) has been obtained, from the ratio between cathode and pixel signals.\\
The detector has been designed for implementation on a small Compton telescope on a CubeSat platform, named \emph{MeVCube}. Given the measured energy and spatial resolution for CdZnTe detector, \emph{MeVCube} sensitivity is comparable to the one achieved by the last generation of large satellites like \emph{COMPTEL} and \emph{INTEGRAL}, as shown in \cite{Lucchetta_MeVCube}. The next steps will include the design and test of new carrier boards for the CdZnTe detectors, with focus on optimization of their noise performance and qualification for space operation. Validation of the temperature dependence of the detector performance will be carried out in a thermal vacuum chamber. The development of a prototype Compton camera employing multiple CdZnTe detectors will allow a full evaluation of the Compton event reconstruction performance.

\appendix
\section{VATA450.3 as a low-power read-out electronics for space operations}
\label{sec:VATA450.3}

\subsection{VATA450.3 overview}
\begin{figure}[htbp]
\centering 
\includegraphics[width=0.95\textwidth]{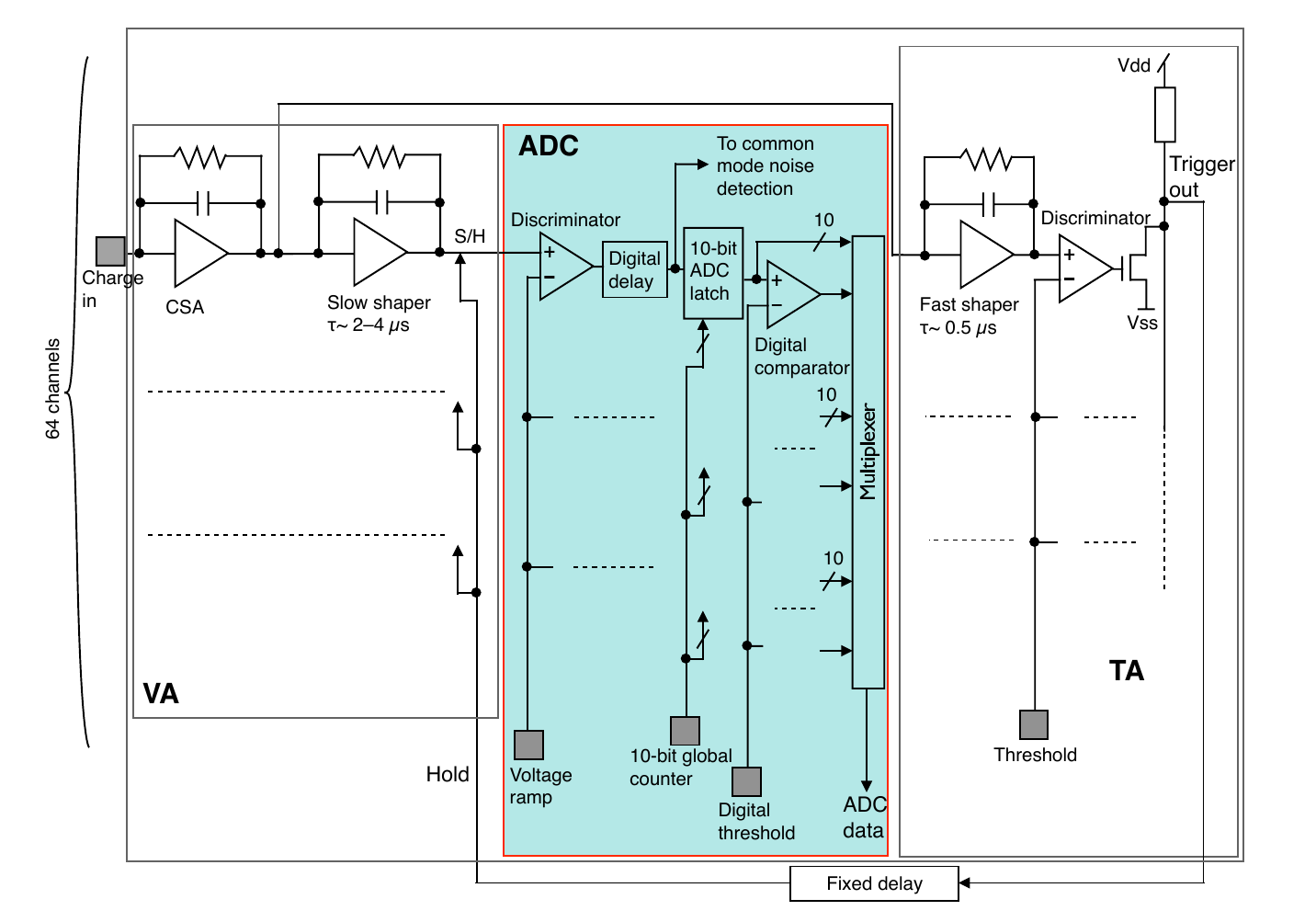}
\caption{\label{fig:VATA_Overview} Circuit diagram of VATA450.3. Picture taken from \cite{Tajima_VATA450}.}
\end{figure}

VATA450.3 is a $64$ channel ASIC developed by \emph{Ideas}, optimized for the front-end readout of CdTe and CdZnTe devices. A schematic representation of VATA450.3's circuit diagram (from \cite{Tajima_VATA450}) is shown in Figure~\ref{fig:VATA_Overview}, while Table~\ref{tab:VATA450_Parameters} summarizes its main parameters and performance.

\begin{table}[htbp]
\centering
\caption{\label{tab:VATA450_Parameters} Summary of main design parameters and performance of VATA450.3, as specified by the manufacturer.}
\begin{center}
\begin{tabular}{ll}
\hline
\hline
\textit{Parameter} & \textit{Value} \\
\hline
\hline
Supplier & \textit{Ideas}\\
\hline
Technology & $0.35\;\mathrm{\mu m}$ CMOS\\
\hline
Chip size & $10000\;\mathrm{\mu m} \times 6500\;\mathrm{\mu m} \times 450\;\mathrm{\mu m}$\\
\hline
Supply voltages & $-2.0\;\mathrm{V} / +1.5\;\mathrm{V}$\\
\hline
Power consumption & $255\;\mathrm{\mu W/channel}$\\
\hline
Max. readout rate & $10\;\mathrm{MHz}$\\
\hline
Shaping time & $\sim 0.6 \;\mathrm{\mu s}$, fast shaper (triggering)\\
	& $\sim 4.0 \;\mathrm{\mu s}$, slow shaper (spectroscopy)\\
\hline
Dynamic range & $-30\;\mathrm{fC} / +16\;\mathrm{fC}$, high gain mode\\
& $-60\;\mathrm{fC} / +50\;\mathrm{fC}$, low gain mode\\
\hline
Non linearity (max.) & $3.2\%$\\
\hline
Equivalent noise charge (ENC) & $45\mathrm{e^-}$ high gain mode, no inputs bonded\\
	& $65\mathrm{e^-}$ low gain mode, no inputs bonded\\
\hline
\hline
\end{tabular}
\end{center}
\end{table}

\begin{figure}[htbp]
\centering
\includegraphics[width=0.95\textwidth]{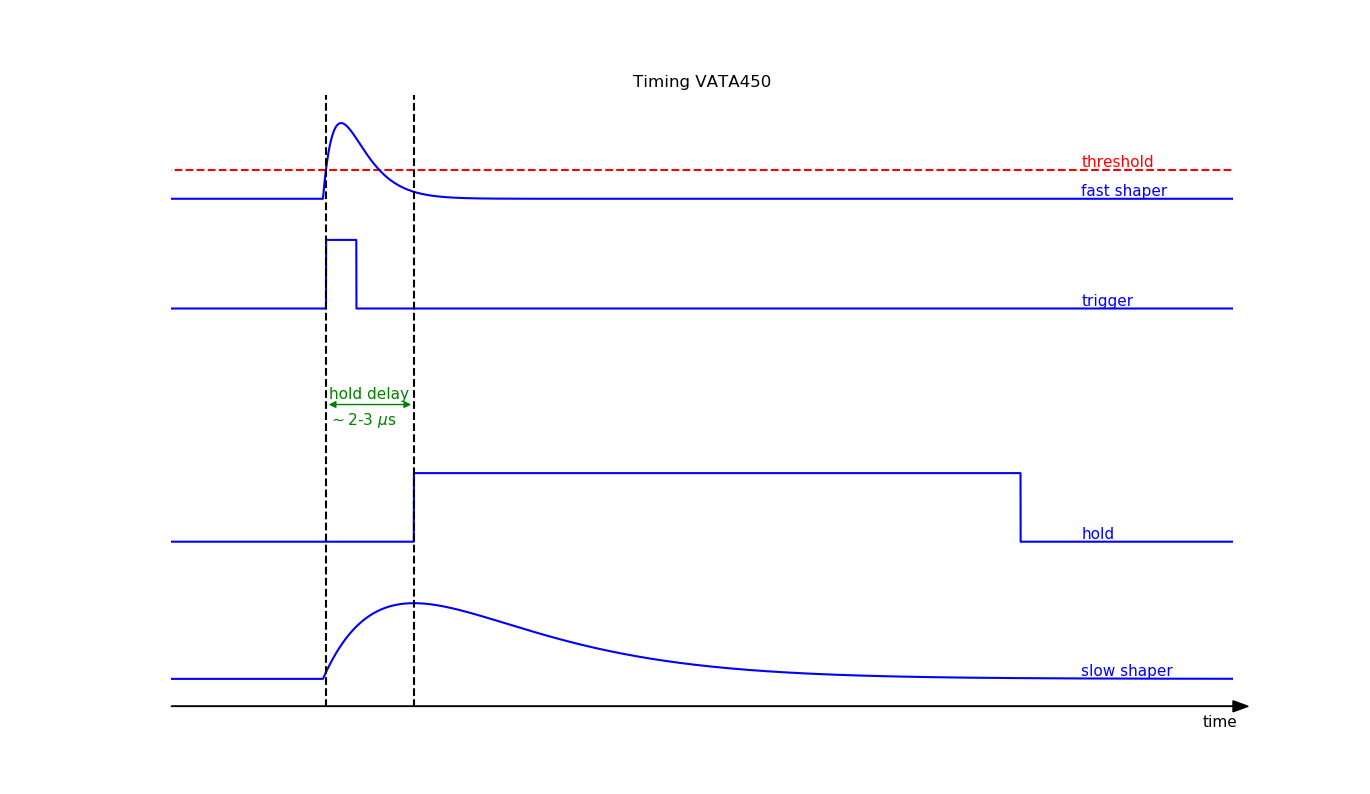}
\caption{\label{fig:VATAprinciple} Sketch illustrating the principle of operation of VATA450.3. The output signal from the charge-sensitive pre-amplifier is shaped by a fast shaper for triggering and slow shaper for spectroscopy. When an adjustable threshold voltage is reached for the fast shaper output, the trigger signal is generated. The hold signal, generated after a fixed delay (\emph{hold delay}), is used to sample the output signal from the slow shaper. The \emph{hold delay} value is chosen in order to probe this signal approximately at its maximum height.} 
\end{figure}

\noindent Each channel of the ASIC implements a charge-sensitive pre-amplifier (CSA) followed by a slow shaper for spectroscopy and a fast shaper for triggering. The fast shaper has a shaping time of $\sim 0.6\;\mathrm{\mu s}$ and is followed by a discriminator to generate the trigger signal (TA component in Figure~\ref{fig:VATA_Overview}).  A second shaper, which generates the output pulse, has a longer shaping time of about $4.0\;\mathrm{\mu s}$. With a sample and hold circuit the pulse height is sampled at the time specified by an external hold signal, produced from the trigger signal with a configurable delay (referred as \emph{hold delay} or \emph{fixed delay}). Figure~\ref{fig:VATAprinciple} illustrates this principle of operation. Each of the $64$ analog signals is then converted to digital values with a $10$-bit Wilkinson-type ADC (Analogue-to-Digital Converter); a multiplexer finally forwards all the $64$ signal from the ADCs to the output data-stream.\\
VATA450.3 can operate with either negative or positive input charges and in two different gain modes (high gain mode and low gain mode). All amplifier inputs are protected against over-voltage and electrostatic discharge; input and output circuits are also designed to allow daisy-chaining of multiple ASICs. 

\subsection{VATA450.3 performance: dynamic range, noise and linearity}
\begin{figure}[htbp]
	\centering
	\subfloat[Dynamic range of VATA450.3 for different settings of the ASIC in the low gain mode and for positive input charge.]{
	\includegraphics[width=0.48\textwidth]{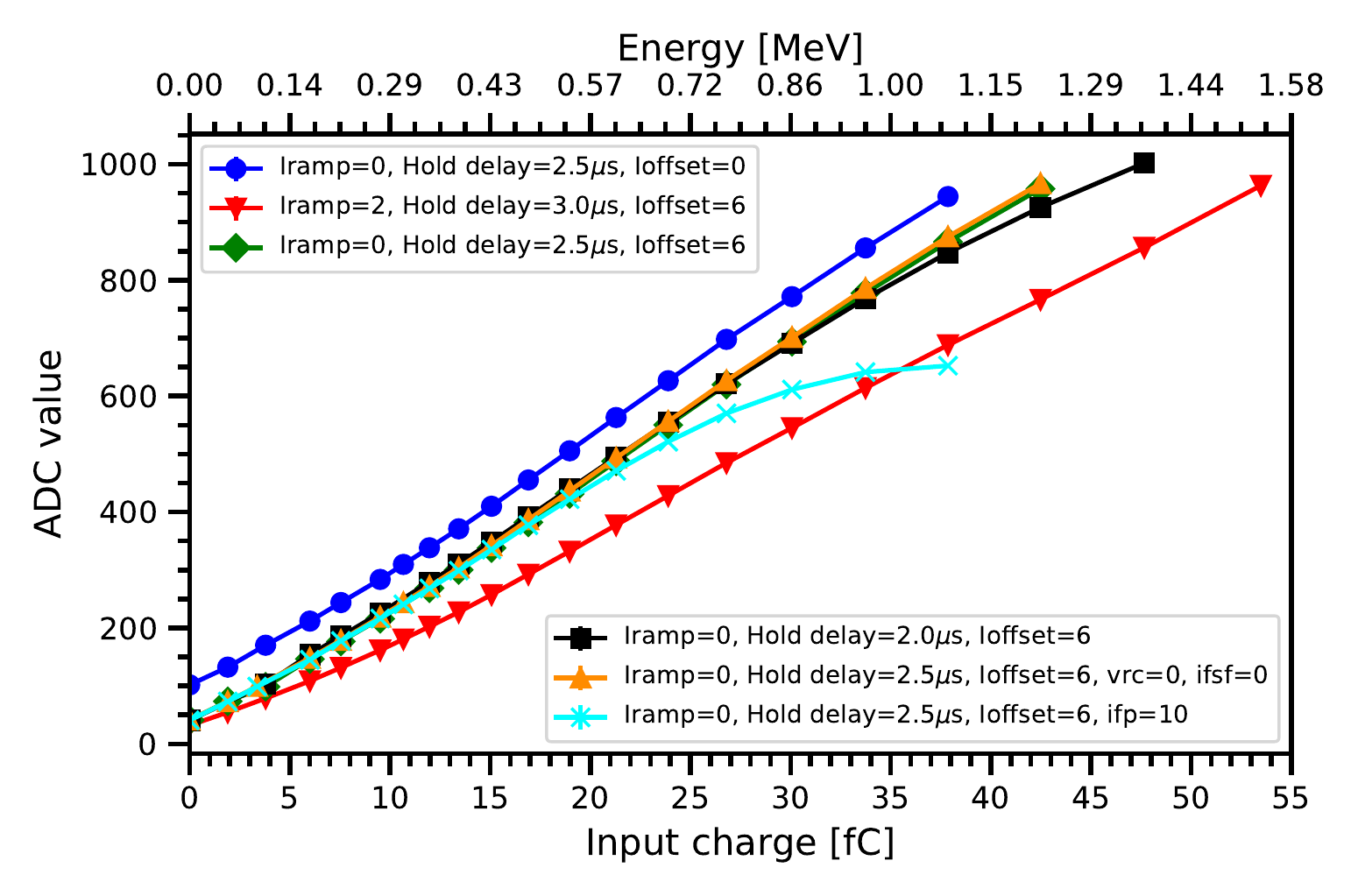}
	\label{fig:VATA_DynamicRangePos}}
	\hspace{3pt}
	\subfloat[Dynamic range of VATA450.3 for different settings of the ASIC in the low gain mode and for negative input charge.]{
	\includegraphics[width=0.48\textwidth]{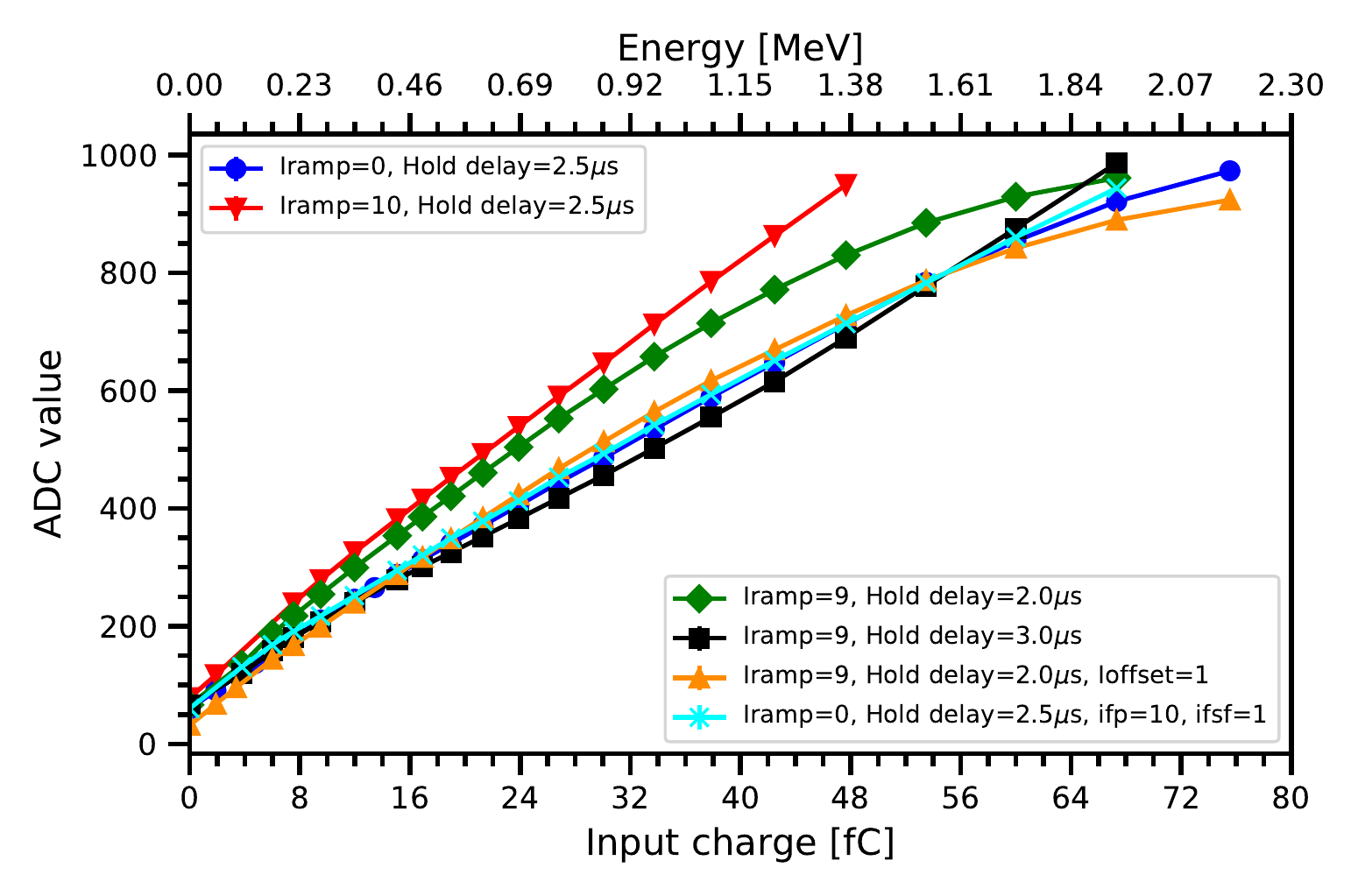}
	\label{fig:VATA_DynamicRangeNeg}}
	\quad
	\subfloat[Noise performance of VATA450.3 for different load capacitors in the low gain mode.]{
	\includegraphics[width=0.48\textwidth]{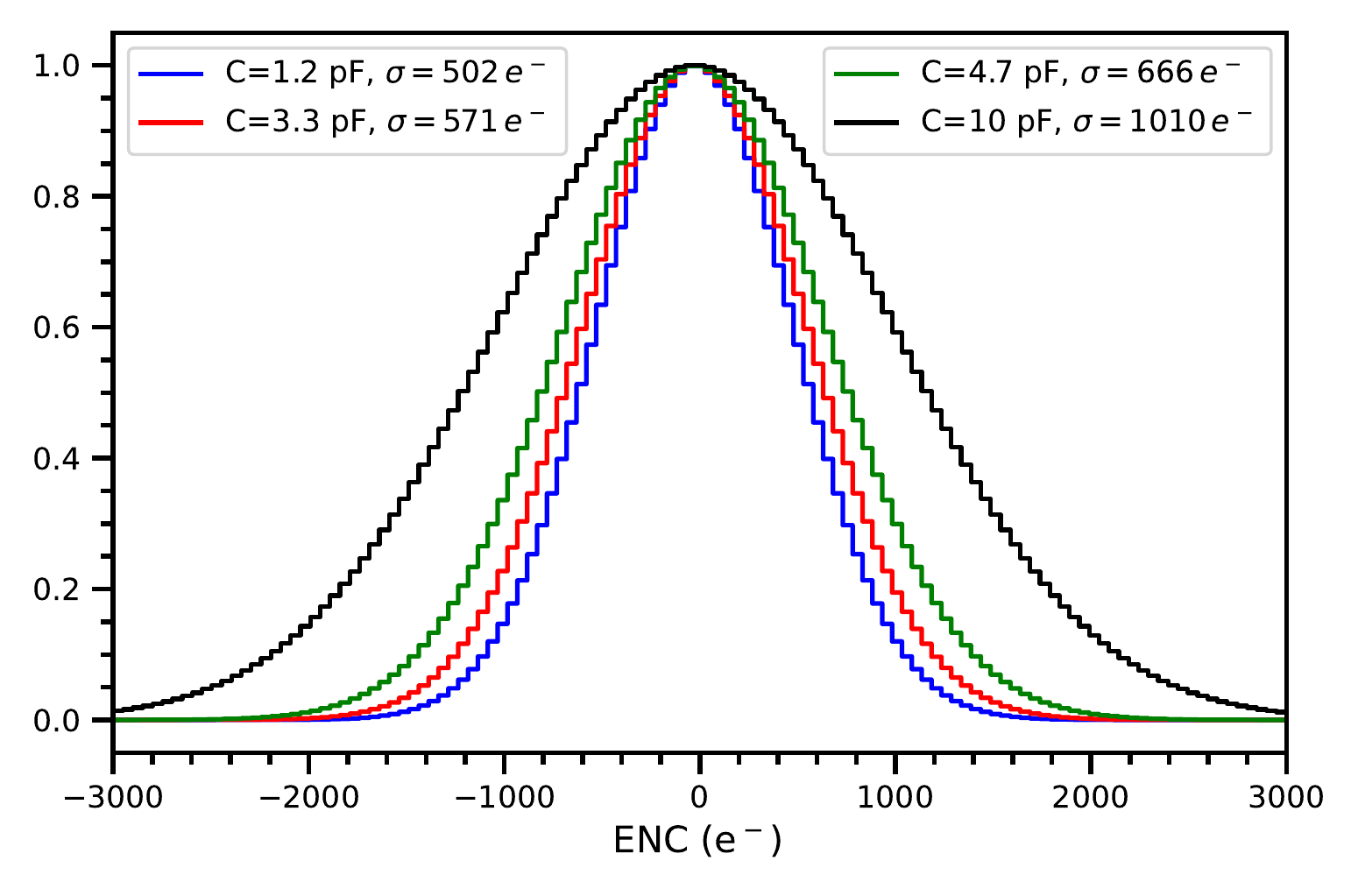}
	\label{fig:VATAnoise}}
	\hspace{3pt}
	\subfloat[Equivalent noise charge (black solid line) and integral non-linearity (red dash-dot line) for all $64$ VATA450.3 channels. Measurements are performed for the specified settings corresponding to the red line in Figure~\ref{fig:VATA_DynamicRangeNeg}.]{
	\includegraphics[width=0.48\textwidth]{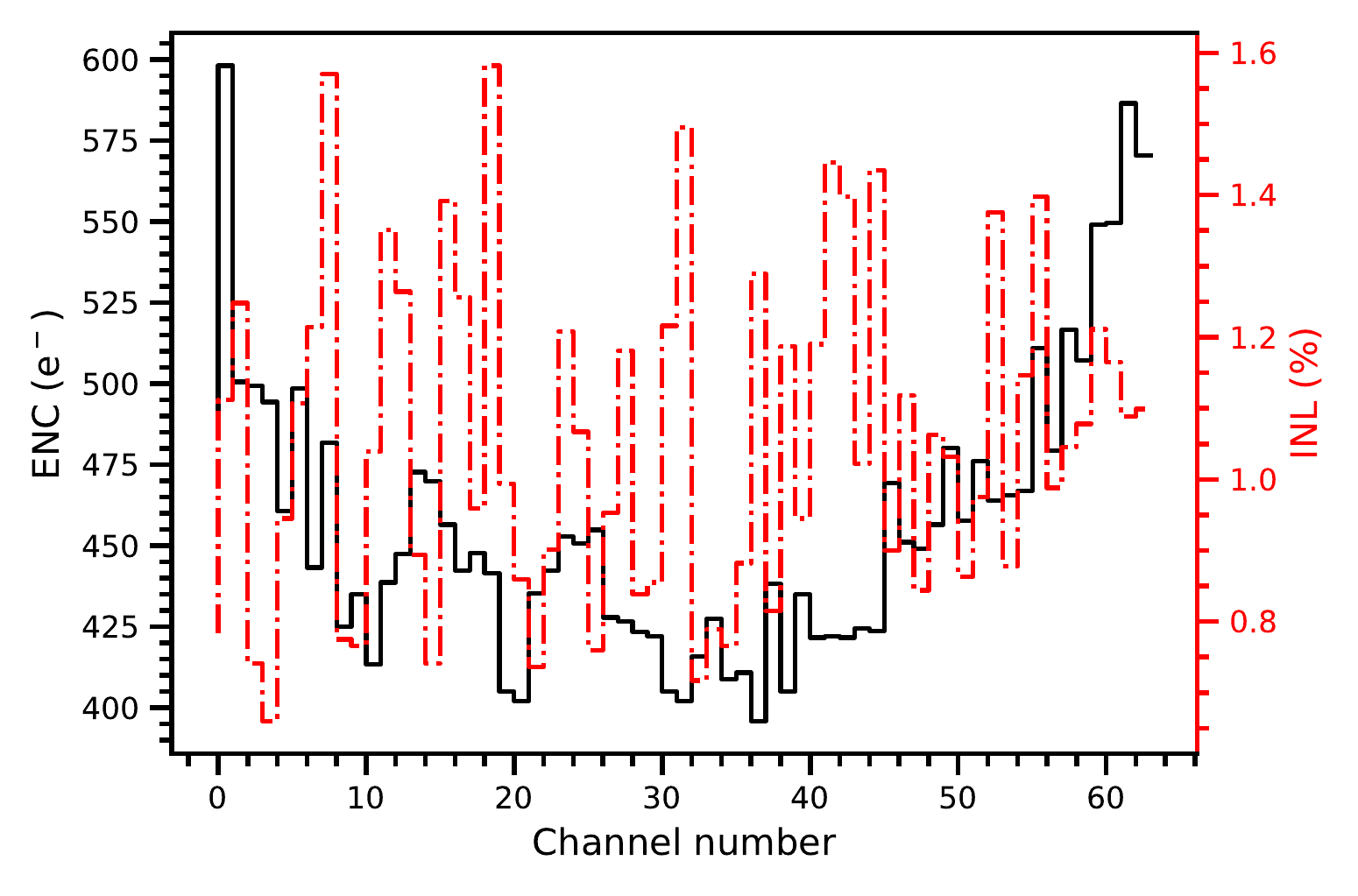}
	\label{fig:AllChannels}}
	\caption{\label{fig:VATAperformance} Main performance of the VATA450.3 ASIC: dynamic range, noise, integral non-linearity.}
\end{figure}

Our tests have been performed on evaluation boards and control software directly provided by \emph{Ideas}. A \emph{Galao} evaluation board, with a Xilinx FPGA\footnote{Field-Programmable Gate Array.} and bias generators, configures the VATA450.3 ASIC for different working modes and controls data read-out and communication with the host computer. The VATA test-board hosts one VATA450.3 ASIC, input connectors for detectors and test points for all key VATA450.3 signals.\\
Input calibration signals can be provided via an internal pulse directly generated by the evaluation board, or from an external pulse generator, connected to the input connectors of the VATA test-board, through different load capacitors. Figure~\ref{fig:VATA_DynamicRangePos} and Figure~\ref{fig:VATA_DynamicRangeNeg} show the dynamic range of VATA450.3 in the low-gain mode. Measurements have been performed with charge of both polarities: positive input charge and negative input charge, respectively. The different curves correspond to different settings of VATA450.3, like the ADC ramp speed (\emph{Iramp}) and offset (\emph{Ioffset}), the \emph{hold delay}, and the bias voltages controlling the feedback MOS in the pre-amplifier and shapers (\emph{ifp}, \emph{ifss}, \emph{ifsf}). The values are set by internal DACs\footnote{Digital to Analogue Converters.} governed via the control register. The figures verify the manufacturer specifications, that VATA450.3 can cover the range up to $+50\;\mathrm{fC}$ for a positive input charge and $-60\;\mathrm{fC}$ for a negative input charge, with an integral non-linearity of just few percents. Since the average electron-hole pair creation energy for CdZnTe is around $4.6\;\mathrm{eV}$, it follows that VATA450.3 can be coupled to the detector in order to measure charge deposits from gamma rays up to $1$ - $2\;\mathrm{MeV}$, suitable for our application. For negative signals, at the expense of linearity, the dynamic range can be extended to $-80\;\mathrm{fC}$.

\begin{figure}[htbp]
\centering
\includegraphics[width=0.75\textwidth]{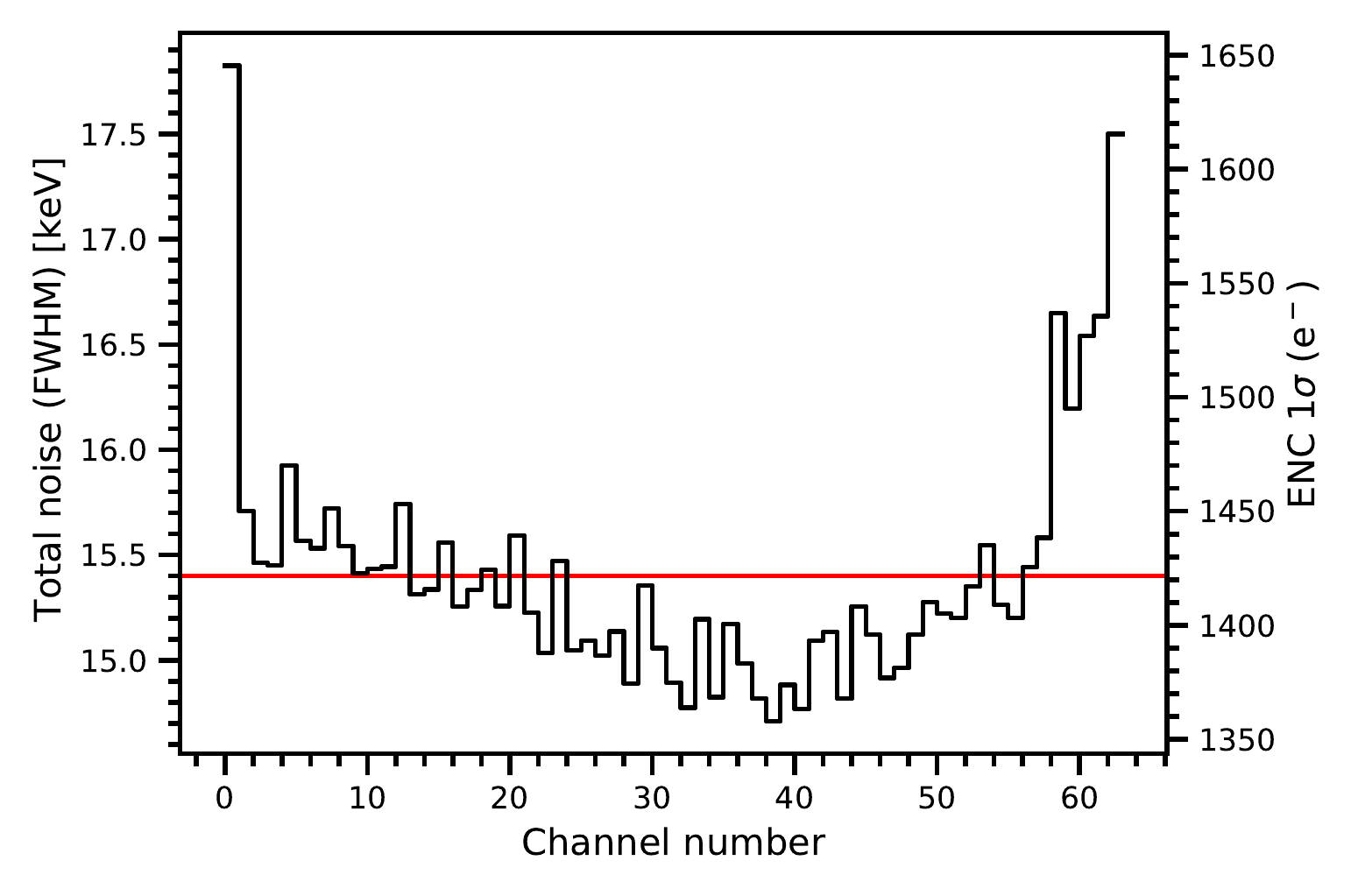}
\caption{\label{fig:NoiseFinalSetUp} Total noise measured from the width of the pedestal distribution in the final set-up, expressed both in $\mathrm{keV}$ as FWHM and in equivalent noise charge ($e^-$). The conversion between $\mathrm{keV}$ and $e^-$ is done taking into account an average electron-hole pair creation energy of $4.6\;\mathrm{eV}$ in CdZnTe; an additional factor of $2.355$ takes into account the passage from $\sigma$ to FWHM. The horizontal red line displays the average for all channels.} 
\end{figure}

\noindent Figure~\ref{fig:VATAnoise} summarizes the noise performance of VATA450.3, expressed in \emph{equivalent noise charge} (ENC). Measurements have been taken for different load capacitors from $1.2\;\mathrm{pF}$ to $10\;\mathrm{pF}$, computing the noise from the width of the calibration pulse (as $1\,\sigma$ unit). A plot of the integral non-linearity\footnote{i.e. the deviation of the ADC values from the best straight line describing the dynamic range of the device.} and the noise (for a $1.2\;\mathrm{pF}$ load capacitor) for all $64$ channels of the ASIC is reported in Figure~\ref{fig:AllChannels}, measured for the settings corresponding to the red line in Figure~\ref{fig:VATA_DynamicRangeNeg}. In the final set-up with the detector attached to the evaluation board, the total noise contribution of the system has been measured from the width of the pedestal distribution: the results for all $63$ pixels are shown in Figure~\ref{fig:NoiseFinalSetUp}.

\section{Steering grid measurements}
\label{sec:SteeringGrid}
In pixelated CdZnTe detectors, charge sharing causes charge loss for events occurring in the gap between pixels, and is one of the factors that degrades the performance of the detector. The presence of a steering grid, surrounding the anode pixel and biased to a slightly negative voltage with respect to the grounded pixels, has shown to improve the performance of CdZnTe detectors \cite{SteeringGrid, ChargeSharing_Kim}: when the steering grid is biased to a slightly negative voltage, electrons are forced to move towards the pixels when approaching the anode surface, reducing charge loss in the pixels gap. Typical values for the steering grid voltage are $\sim \, -50\;\mathrm{V}$ \cite{SteeringGrid}: lower values do not lead to significant improvements in collection efficiency, while higher values result in an additional surface current and increased noise. The performance degradation due to charge sharing effect, may vary according to several factors of the considered device, such as pixel size, width of the gap between pixels, size of the steering grid, as well as shaping performance and noise of the read-out electronics.

\begin{figure}[htbp]
	\centering
	\subfloat[Pixel G4.]{
	\includegraphics[width=0.49\textwidth]{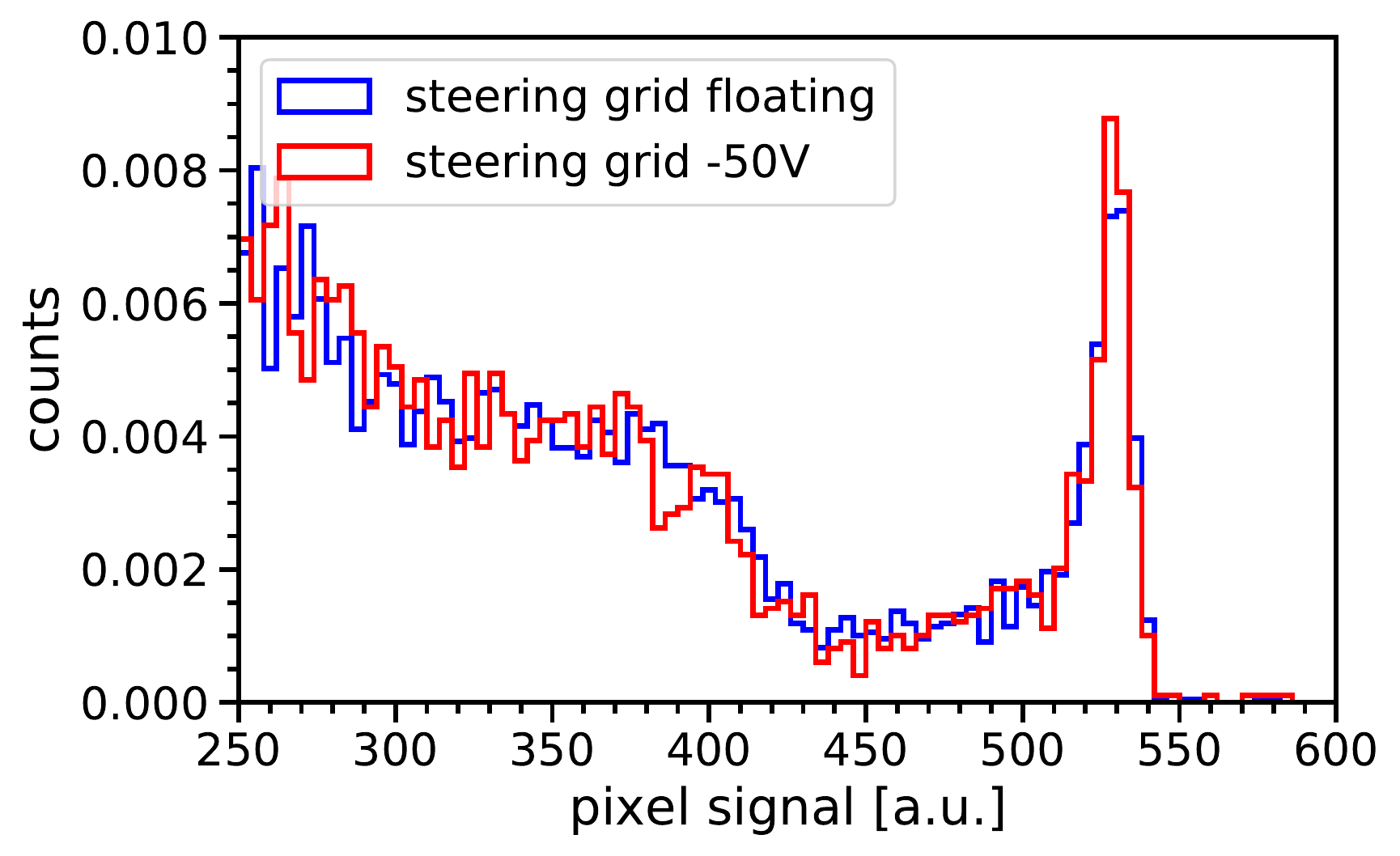}}
	\vspace{\floatsep}
	\subfloat[Pixel H8.]{
	\includegraphics[width=0.49\textwidth]{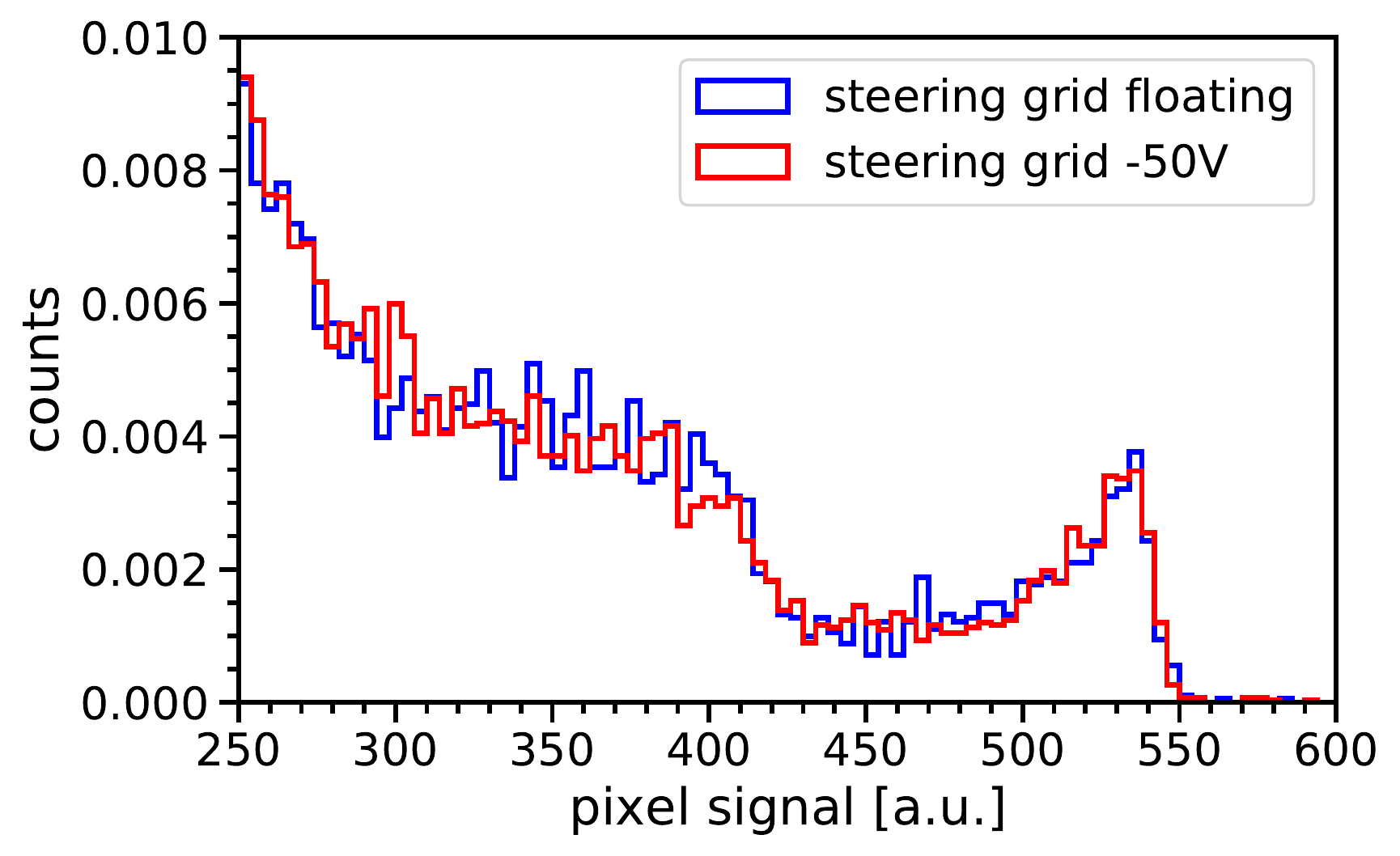}}
	\caption{\label{fig:Spectra_SteeringGrid} Comparison of the energy spectra obtained from measurements with the Cs-$137$ radioactive source for two selected pixels, with the steering grid floating and the steering grid biased at $-50\;\mathrm{V}$. The histograms have been normalized to allow an easier comparison of the respective shapes.}
\end{figure}

\noindent In our set-up two sets of measurements were performed, in order to evaluate the effect of the steering grid: one with the steering grid floating and one with the steering grid biased at $-50\;\mathrm{V}$. The detector was tested under irradiation with a Cs-$137$ source. In Figure~\ref{fig:Spectra_SteeringGrid} the spectra obtained with the steering grid floating and the steering grid biased at $-50\;\mathrm{V}$ are compared. No overall improvement in the energy resolution is measured before and after depth-of-interaction correction. In order to quantify the improvement in efficiency due to the steering grid biasing, the detection rates have been considered. The detection rate is determined by counting the number of events in the photo-peak $N$, normalized to the observation time $t$. Then, the relative improvement in efficiency is computed as the difference between the detection rate measured with the steering grid biased at $-50\;\mathrm{V}$ and the detection rate measured with the steering grid floating:
\begin{equation}
\Delta \epsilon = \frac{\frac{N_{-50\mathrm{V}}}{t_{-50\mathrm{V}}} - \frac{N_{\mathrm{floating}}}{t_{\mathrm{floating}}}}{\frac{N_{\mathrm{floating}}}{t_{\mathrm{floating}}}}\, .
\end{equation}

\begin{figure}[htbp]
	\centering
	\includegraphics[width=0.70\textwidth]{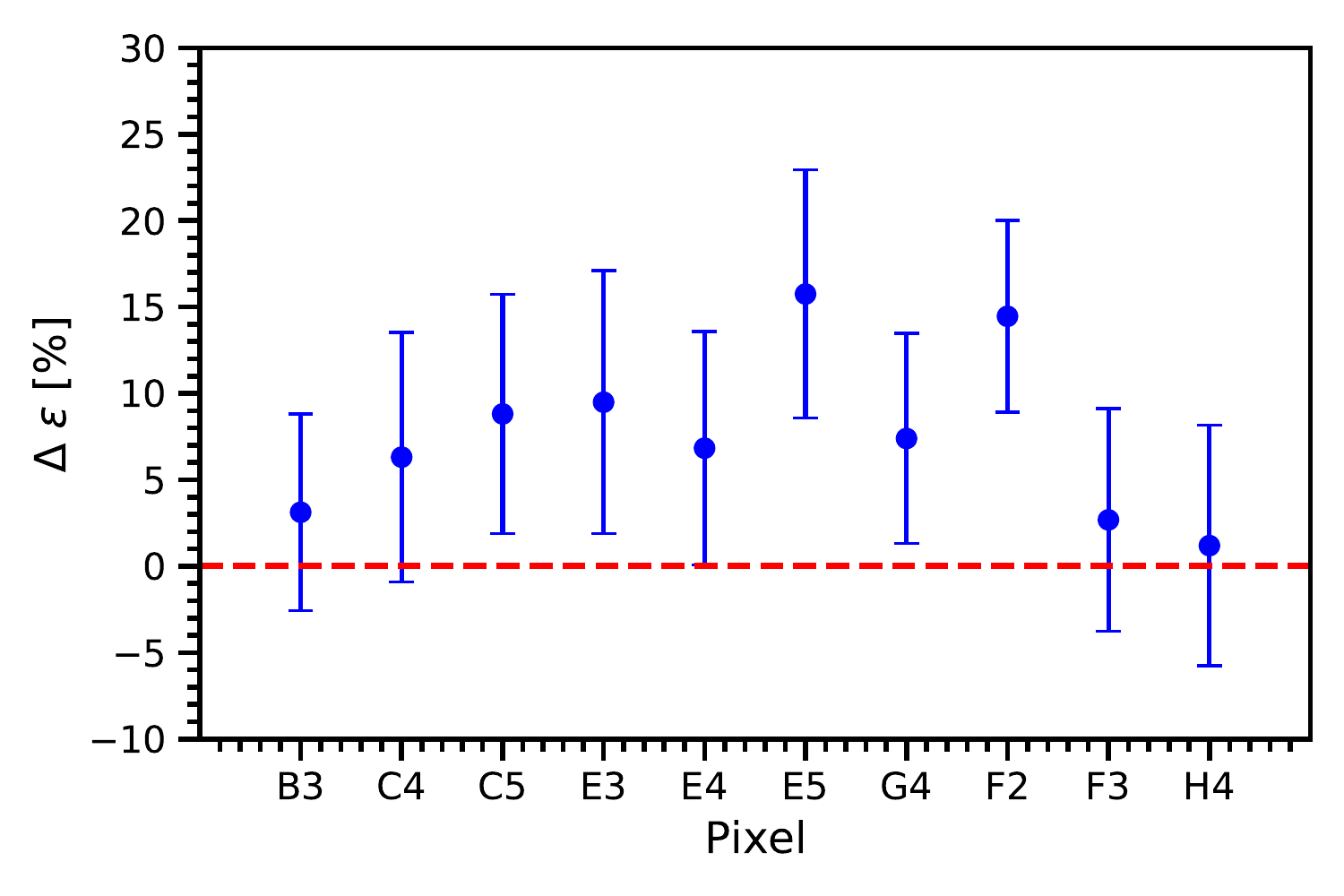}
	\caption{\label{fig:SteeringGrid} Improvement of the detection efficiency for $10$ selected pixels when biasing the steering grid.}
\end{figure}

\noindent The results of our analysis, for $10$ selected pixels, are shown in Figure~\ref{fig:SteeringGrid}: as it can be observed the biasing of the steering grid results in a slightly improved detection efficiency. At the same time, however, the steering grid biasing introduces some design complications, such as an additional power supply and filtering circuit to provide the required negative voltage, and the loss of an active pixel due to the space left for the electrical contact of the steering grid. Therefore, for our application, the improvement derived from a biased steering election is thought to be too small to justify the additional system complexity deriving from it.


\acknowledgments

This publication is funded by the Deutsche Forschungsgemeinschaft (DFG, German Research Foundation) - 491245950.

\end{document}